\documentclass[aps,prd,preprint,superscriptaddress,tightenlines,nofootinbib]{revtex4}

\usepackage{amsfonts,amsmath,amsthm,amssymb}
\usepackage{mathrsfs}
\usepackage{bm}
\usepackage{color}
\usepackage{epsfig}

\usepackage[usenames,dvipsnames]{xcolor}
\definecolor{orange}{cmyk}{0,0.5,1,0}
\definecolor{rossoCP3}{cmyk}{0,.88,.77,.40}
\definecolor{graa}{rgb}{0.8,0.8,0.8}
\definecolor{blaa}{rgb}{0.2,0.2,0.6}
\def\CB{{\cal B}}

\newcommand{\PRE}[1]{{#1}}   
\newcommand{\met} {\not\!\! E_T}

\newcommand{\beq}{\begin{equation}}
\newcommand{\eeq}{\end{equation}}
\newcommand{\bea}{\begin{flushleft} \begin{eqnarray}}
\newcommand{\eea}{\end{eqnarray}\end{flushleft}}

\newcommand{\postscript}[2]{\setlength{\epsfxsize}{#2\hsize}
   \centerline{\epsfbox{#1}}}
\newcommand{\comment}[1]{}

\newcommand{\ci}[1]{}

\newcommand{\lsb}{\left[}
\newcommand{\rsb}{\right]}

\newcommand{\p}{\partial}
\newcommand{\pd}{\partial}

\newcommand{\ba}{\begin{eqnarray}}
\newcommand{\ea}{\end{eqnarray}}
\newcommand{\be}{\begin{equation}}
\newcommand{\ee}{\end{equation}}
\newcommand{\bay}[1]{\left(\begin{array}{#1}}
\newcommand{\eay}{\end{array}\right)}

\def\met{\mbox{${\hbox{$E$\kern-0.6em\lower-.1ex\hbox{/}}}_T$}} 

\def\CF{{\cal F}}

\def\CM{{\cal M}}

\newcommand{\beqa}{\begin{eqnarray}}
\newcommand{\eeqa}{\end{eqnarray}}

\newcommand{\la}{\langle}
\newcommand{\ra}{\rangle}
\newcommand{\lpa}{\left(}
\newcommand{\rpa}{\right)}

\begin{document}

\title{\PRE{\vspace*{0.9in}} \color{rossoCP3}{ 
    Weinberg's Higgs portal confronting  recent LUX and LHC results 
    together with upper limits on  $\bm{B^+}$ and $\bm{K^+}$ decay into invisibles}
\PRE{\vspace*{0.1in}} }

\author{Luis A.~Anchordoqui}
\affiliation{Department of Physics,\\
University of Wisconsin-Milwaukee,
 Milwaukee, WI 53201, USA
\PRE{\vspace*{.05in}}
}

\author{Peter B.~Denton}
\affiliation{Department of Physics and Astronomy,\\
Vanderbilt University, Nashville TN 37235, USA
\PRE{\vspace*{.1in}}
}

\author{Haim~\nolinebreak Goldberg}
\affiliation{Department of Physics,\\
Northeastern University, Boston, MA 02115, USA
\PRE{\vspace*{.1in}}
}

\author{Thomas~\nolinebreak C.~\nolinebreak Paul}
\affiliation{Department of Physics,\\
University of Wisconsin-Milwaukee,
 Milwaukee, WI 53201, USA
\PRE{\vspace*{.05in}}
}

\affiliation{Department of Physics,\\
Northeastern University, Boston, MA 02115, USA
\PRE{\vspace*{.1in}}
}

\author{Luiz~\nolinebreak H.~\nolinebreak M.~\nolinebreak da~\nolinebreak
  Silva}
\affiliation{Department of Physics,\\
University of Wisconsin-Milwaukee,
 Milwaukee, WI 53201, USA
\PRE{\vspace*{.05in}}
}

\author{Brian J. Vlcek}
\affiliation{Department of Physics,\\
University of Wisconsin-Milwaukee,
 Milwaukee, WI 53201, USA
\PRE{\vspace*{.05in}}
}

\author{Thomas J. Weiler}
\affiliation{Department of Physics and Astronomy,\\
Vanderbilt University, Nashville TN 37235, USA
\PRE{\vspace*{.1in}}
}

\date{December 2013}

\begin{abstract}
 \PRE{\vspace*{.1in}} 
\noindent We discuss a number of experimental constraints on
  Weinberg's Higgs portal model. In this framework, the standard model
  (SM) particle spectrum is extended to include one complex scalar
  field $S$ and one Dirac fermion $\psi$.  These new fields are
  singlets under the SM gauge group and are charged under a global
  $U(1)$ symmetry. Breaking of this $U(1)$ symmetry results in a
  massless Goldstone boson $\alpha$ and a massive $CP$-even scalar
  $r$, and splits the Dirac fermion into two new mass-eigenstates
  $\psi_\pm$, corresponding to Majorana fermions. The interest on such
  a minimal SM extension is twofold. On the one hand, if the Goldstone
  bosons are in thermal equilibrium with SM particles until the era of
  muon annihilation their contribution to the effective number of
  neutrino species can explain the hints from cosmological
  observations of extra relativistic degrees of freedom at the epoch
  of last scattering. On the other hand, the lightest Majorana fermion
  $\psi_-$ provides a plausible dark matter candidate. Mixing of $r$
  with the Higgs doublet $\phi$ is characterized by the mass of hidden
  scalar $m_h$ and the mixing angle $\theta$. We constrain this
  parameter space using a variety of experimental data, including
  heavy meson decays with missing energy, the invisible Higgs width,
  and direct dark matter searches.  We show that different
  experimental results compress the allowed parameter space in
  complementary ways, covering a large range of $\psi_-$ masses
  ($5~{\rm GeV} \alt m_- \alt 100~{\rm GeV}$).  Though current results
  narrow the parameter space significantly (for the mass range of
  interest, $\theta \alt 10^{-3}$ to $10^{-4}$), there is still room
  for discovery ($\alpha$ decoupling at the muon annihilation era
  requires $\theta \agt 10^{-5}$ to $10^{-4}$).  In the near future,
  measurements from ATLAS, CMS, LHCb, NA62, XENON1T, LUX, and CDMSlite will
  probe nearly the full parameter space.
\end{abstract}
\pacs{xxxx}
\maketitle

\section{Introduction}

With the success of the Large Hadron Collider (LHC) at CERN, a new era
of discovery has begun. The $SU(3)_C \times SU(2)_L \times U(1)_Y$
standard model (SM) of electroweak and strong interactions has once
again endured intensive scrutiny, with a dataset corresponding to an
integrated luminosity of $\approx 20~{\rm fb}^{-1}$ of $pp$ collisions
at $\sqrt{s} = 8~{\rm TeV}$.  Most spectacularly, the recent
discovery~\cite{Aad:2012tfa,Chatrchyan:2012ufa} of a particle which
seems to be the SM Higgs has possibly plugged the final remaining
experimental hole in the SM, cementing the theory further.  The LHC8
data have not yet turned up any evidence of physics beyond the
SM~\cite{Altarelli:2013lla}.

Despite the resilience of the SM, it seems clear that there is more to
the story. The concordance model of cosmology -- a flat expanding
universe comprising 5\% baryons, 20\% dark matter, and 75\% dark
energy -- is achieving an ever-firmer footing thanks to observations
of the Supernova Cosmology
Project~\cite{Perlmutter:1998np,Knop:2003iy,Kowalski:2008ez}, the
Supernova Search Team~\cite{Riess:1998cb,Riess:2001gk,Tonry:2003zg},
the Wilkinson Microwave Anisotropy Probe
(WMAP)~\cite{Spergel:2003cb,Komatsu:2010fb,Hinshaw:2012fq}, the Hubble
Space Telescope~\cite{Riess:2009pu,Riess:2011yx}, the Sloan Digital
Sky Survey
(SDSS)~\cite{Abazajian:2003jy,Tegmark:2003ud,Abazajian:2008wr,Percival:2009xn},
and the Planck spacecraft~\cite{Ade:2013lta}.  These observations are
propounding evidence that a description of the physics of the early
universe, and thus the particle physics interactions at sub-fermi
distances, will require new theoretical concepts which transcend the
SM.

The existence of dark matter (DM), has been solidified by multiple
astrophysical observations~\cite{Feng:2010gw}.  Weakly interacting
massive particles (WIMPs) are among the best motivated 
candidates~\cite{Steigman:1984ac}. If stable particles with mass and
annihilation cross section set by the weak scale exist, they would be
produced and annihilate in thermal equilibrium in the early Universe.
As the Universe expands, these particles fall out of equilibrium and
their number density is frozen in. A typical weak scale interaction
rate yields a thermally-averaged WIMP annihilation cross section,
$\langle \sigma v_M \rangle \sim 10^{-9}~{\rm GeV}^{-2}$, which naturally produces a WIMP relic density
$h^2 \Omega_{\rm WIMP} \sim 10^{-10}~{\rm GeV}^{-2}/\langle \sigma v_M \rangle$~\cite{Lee:1977ua,Dicus:1977nn,Kolb:1985nn,Scherrer:1985zt,Steigman:2012nb} consistent with
the measured DM abundance  $h^2 \Omega_{\rm DM} = 0.111(6)$~\cite{Beringer:1900zz}, thus making WIMPs promising
candidates of DM.\footnote{Throughout this work we adopt the usual convention of
  writing the Hubble constant at the present day as $H_0 = 100 \
  h~{\rm km} \ {\rm s}^{-1} \ {\rm Mpc}^{-1}$. For $t= {\rm today}$,
  the various energy densities are expressed in units of the critical
  density $\rho_c$; {\it e.g.}, the DM density $\Omega_{\rm DM}
  \equiv \rho_{\rm DM}/\rho_c$.} 

Since WIMPs are subject to the weak interaction, it is possible to
search for them via direct detection experiments, $\gamma$-ray
observatories, neutrino telescopes, and particle colliders.  The first
direct detection experiment to claim evidence for DM was
DAMA/LIBRA~\cite{{Bernabei:1998fta}}, which has recorded an annual
modulation in nuclear recoil event rate at the 8.9$\sigma$
level~\cite{Bernabei:2010mq}.  This modulation can be interpreted as a
consequence of the change in the relative motion of the detector
through the sea of DM as the Earth rotates around the
Sun~\cite{Drukier:1986tm, Freese:1987wu}.  Other direct detection
experiments have provided supporting evidence for WIMP interactions,
including CRESST~\cite{Angloher:2011uu},
CoGeNT~\cite{Aalseth:2010vx,Aalseth:2011wp,Aalseth:2012if}, and most
recently the CDMS II~\cite{Agnese:2013rvf} experiment. Interestingly,
all of these observations favor a light WIMP, with mass $\sim 10~{\rm
  GeV}$ and an interaction with protons via spin-independent elastic
scattering with a cross-section $\sim 10^{-41}~{\rm cm}^2$. In
contrast, the XENON-10~\cite{Angle:2011th} and
XENON-100~\cite{Aprile:2011hi} DM experiments have reported limits
which exclude the mass and cross-section regime favored by CoGeNT,
CRESST and CDMS II.

A variety of models were employed to reconcile hints of the signals
mentioned above with the exclusion from XENON-10 and XENON-100.
However, tension has increased even further after recent CDMSlite (for
CDMS Low Ionization Threshold Experiment)~\cite{Agnese:2013lua} and
LUX~\cite{Akerib:2013tjd} results.  At this point only the xenophobic
isospin violating dark
matter~\cite{Kurylov:2003ra,Giuliani:2005my,Feng:2011vu,Frandsen:2013cna,Feng:2013vod},
with a neutron to proton coupling ratio of $-0.7$ allows any overlap
with the 68\% favored contour of CDMS
II~\cite{Gresham:2013mua,DelNobile:2013gba,Cirigliano:2013zta}. Favored regions of all other experiments remain excluded.

Adding to the story, the most recent data from
WMAP~\cite{Komatsu:2010fb,Hinshaw:2012fq}, the Atacama Cosmology
Telescope ~\cite{Dunkley:2010ge}, and the South Pole
Telescope~\cite{Keisler:2011aw} have hinted at a higher value of the
fractional energy density in relativistic species than previously
estimated~\cite{Hamann:2011hu,Benetti:2013wla}.
Furthermore,  recent estimates of light-element abundances probing
Big Bang nucleosynthesis  also suggest additional relativistic degrees of freedon in
the early Universe~\cite{Steigman:2012ve}.

The energy density stored in relativistic species is customarily given
in terms of the number of “equivalent” light neutrino species,
\begin{equation}
N_{\rm eff} \simeq \frac{8}{7} {\sum_{\rm b}}' \frac{g_{\rm b}}{2}
\left(\frac{T_{\rm b}}{T_\nu} \right)^4 + {\sum_{\rm f}}' \frac{g_{\rm f}}{2}
  \left(\frac{T_{\rm f}}{T_\nu} \right)^4 \,,
\end{equation}
where $g_{\rm b\, (f)}$ are the number of boson (fermion) helicity
states,   $T_{\rm b \, (f )}$ are the  temperatures of the various species, 
and the primes indicate that electrons and photons are excluded from
the sums~\cite{Steigman:1977kc}.  The normalization of $N_{\rm eff}$
is such that it gives $N_{\rm eff} = 3$ for three families of massless
left-handed SM neutrinos, with  temperature $T_\nu$.

The latest chapter in the story is courtesy of the Planck spacecraft.
Unexpectedly, the best multi-parameter fit of Planck data yields a
Hubble constant $h= 0.674 \pm 0.012$~\cite{Ade:2013lta}, a result
which deviates by more than 2$\sigma$ from the value obtained with the
Hubble Space Telescope, $h = 0.738 \pm
0.024$~\cite{Riess:2011yx}.  The impact of the
Planck $h$ estimate is particularly important in the determination of
$N_{\rm eff}$. Combining observations of the cosmic microwave background (CMB)  with data from baryon
acoustic oscillations, the Planck Collaboration reported $N_{\rm eff}
= 3.30 \pm 0.27$~\cite{Ade:2013lta}. However, a combination of the space telescope
measurement $h =  0.738 \pm 0.024$ with the Planck CMB data gives $N_{\rm eff}
=3.62 \pm 0.25$, which suggests new neutrino-like physics (at around
the $2.3 \sigma$ level)~\cite{Ade:2013lta}.

As alluded to already, beyond SM physics may be required to resolve
these tensions. The Higgs sector could provide the promising territory
to introduce new physics, as the couplings are the least
experimentally constrained at present.  Perhaps the most direct
example is the Higgs portal, which connects the SM Higgs to a scalar
field in a hidden sector by an elementary quartic
interaction~\cite{Schabinger:2005ei,Patt:2006fw,Barger:2007im,Barger:2008jx}. One
realization of this concept has been introduced recently by Weinberg
specifically to address the apparent inconsistencies between the
cosmological and astrophysical measurements discussed
above~\cite{Weinberg:2013kea}.  In this framework the SM is extended
by one complex scalar field $S$ and one Dirac fermion field
$\psi$. The new fields are singlets under the SM gauge group and are
charged under a global $U(1)_W$ symmetry, namely: $U(1)_W (\psi) = 1$
and $U(1)_W(S) = 2$. Of course, all the SM fields transform trivially
under the global symmetry.  The spontaneous breaking of this global
symmetry gives rise to a massless Goldstone boson and a $CP$-even
scalar, and splits the Dirac fermion into two new mass-eigenstates 
$\psi_\pm$, corresponding to Majorana fermions. If the Goldstone bosons
are in thermal equilibrium with the SM particles until the era of muon
annihilation, then they can contribute to the effective number of
neutrino species.  Furthermore the symmetry breaking leads naturally
to a dark matter candidate.  Fields with an even (odd) charge under
the global $U(1)$ symmetry will acquire, after symmetry breaking, an
even (odd) discrete charge under a $Z_2$ discrete symmetry.  While the
SM particles are all even under $Z_2$, the Majorana fermions
$\psi_\pm$ are odd.  The lightest particle with odd charge, $\psi_-$,
will be absolutely stable, and thus a plausible dark matter candidate.

In this paper we explore the plausible parameter space of this model.
The outline is as follows.  In Sec.~\ref{sec2} we descibe the main
characteristics of Weinberg's Higgs portal model.  In Sec.~\ref{sec3}
we study the contributions to the effective number of neutrino species
in the plane spanned by the mixing angle between scalars in the
visible and hidden sectors and the mass of the $CP$-even scalar.  In
Sec.~\ref{sec4} we use collider data to constrain this parameter space
via processes involving the hidden scalar. In Sec.~\ref{sec5} we use
data from direct dark matter searches to further constrain the
parameter space via the hidden fermions. In Sec.~\ref{sec6} we explore
the impact of LHC measurements of the invisible width of the Higgs on
the same parameter space. In Sec.~\ref{sec7} we gather our
conclusions.

\section{Weinberg's Higgs Portal Model}
\label{sec2}

The Higgs portal couples a complex singlet field $S$ to the SM doublet
$\Phi$, through which the singlet field interacts with the SM. The
renormalizable Lagrangian density of the model is
\begin{eqnarray}
\mathscr{L}  =   \pd_\mu S^\dagger \ \partial^\mu S  +  \mu^2 \ S^\dagger \ S  -  \lambda
(S^\dagger \ S)^2 -   \ g_\theta \ (S^\dagger \ S) (\Phi^\dagger \ \Phi)
+ \mathscr{L}_{\rm SM}  \,,
\label{eq:one}
\end{eqnarray}
where $\mu$, $\lambda$, and $g_\theta$ are constants and $\mathscr{L}_{\rm SM}$ is the usual SM
Lagrangian. The Higgs sector in $\mathscr{L_{\rm SM}}$ is given by
\begin{equation}
\mathscr{L}_{\rm SM} \supset (D_\mu \Phi)^\dagger \ (D^\mu \Phi) + \mu^2_{\rm
  SM} \Phi^\dagger \Phi - \lambda_{\rm SM} (\Phi^\dagger \Phi)^2 \, .
\end{equation}
Following Weinberg, we write $S$ in terms of two real fields (its massive
radial component and a massless Goldstone boson). The radial field
develops a VEV $\langle r \rangle$ about which the field $S$ is
expanded
\begin{equation}
S = \frac{1}{\sqrt{2}} \left(\langle r \rangle + r(x) \right) \ e^{i\, 2
    \alpha(x)} \, .
\end{equation}
The phase of $S$
is adjusted to make  $\langle \alpha (x) \rangle = 0$. In the unitary
gauge the Higgs doublet is expanded around the VEV as
\begin{equation}
\Phi(x) = \frac{1}{\sqrt{2}} \left( \begin{array}{c} 0 \\ \langle \phi
      \rangle + \phi(x) \end{array} \right) ,
\end{equation} 
with $\langle \phi \rangle = 246~{\rm GeV}$. The fields $\phi$ and
$r$, under the influence of the $g_\theta$--term, mix and become two physical
massive Higgs fields~\cite{Anchordoqui:2013pta} 
\begin{equation}
\left(\begin{array}{c} h \\ H \end{array} \right) = \left( \begin{array}{cc} 
\cos \theta & - \sin \theta \\ \sin \theta & \phantom{-} \cos \theta \end{array} \right) 
\left(\begin{array}{c} r \\ \phi \end{array} \right) \,
\end{equation}
with masses 
\begin{equation}
m_h = \lambda \, \langle r \rangle^2 + \lambda_{\rm SM} \, \langle \phi \rangle^2 - \sqrt{\left(\lambda_{\rm SM} \, \langle \phi \rangle^2 - \lambda \, \langle r \rangle^2 \right)^2 + g_\theta^2 \, \langle r \rangle^2 \, \langle \phi \rangle^2} 
\end{equation}
and
\begin{equation}
m_H = \lambda \, \langle r \rangle^2 + \lambda_{\rm SM} \, \langle \phi \rangle^2 + \sqrt{\left(\lambda_{\rm SM} \,\langle \phi \rangle^2 - \lambda \, \langle r \rangle^2\right)^2 + g_\theta^2 \, \langle r \rangle^2 \, \langle \phi \rangle^2} \,,
\end{equation}
and mixing angle
\begin{equation}
\tan 2 \theta = \frac{g_\theta \, \langle r \rangle \, \langle \phi \rangle}{\lambda_{\rm SM} \langle \phi \rangle^2 - \lambda \langle  r \rangle^2} \, .
\label{teta}
\end{equation}
The small $\theta$ limit leads to the usual SM phenomenology with an
isolated hidden sector. 

Adding in the dark matter sector requires at least one Dirac field
\beq
\mathscr{L}_\psi = i \bar{\psi}\gamma \cdot \pd \psi - m_\psi \bar{\psi} \psi 
- \frac{f}{\sqrt{2}} \bar{\psi^c} \psi \, S^\dagger  -
\frac{f^*}{\sqrt{2}} \bar{\psi} \psi^c \, S \,  .
\label{eq:fortyone}
\eeq 
As advanced in the Introduction,
we assign to the hidden fermion a charge $U(1)_W (\psi) = 1$, so that the Lagrangian is
invariant under the global transformation $e^{i W \alpha}$. Treating 
the transformation as local allows us to express $\psi$ as \beq
\psi(x) = \psi'(x) e^{i \alpha(x)}.
\label{eq:fortytwo}
\eeq
Once the radial field achieves a VEV we
can expand the dark matter sector to get \beqa \mathscr{L}_\psi &=&
\frac{i}{2}\left(\bar{\psi}'\gamma \cdot \pd \psi' + \bar{\psi'}^{c}
  \gamma \cdot \pd \psi^{c'} \right), \nonumber
\\
&-& \frac{m_\psi}{2} \left( \bar{\psi}' \psi' + \bar{\psi'}^{c}
  {\psi'}^{c} \right)-\frac{f \langle r \rangle}{2} \bar{\psi'}^{c} \psi' - \frac{f
  \langle r \rangle}{2} \bar{\psi}' {\psi'}^{c} , \nonumber
\\
&-& \frac{1}{2} (\bar{\psi}' \gamma \psi' - \bar{\psi'}^{c} \gamma
{\psi '}^{c} ) \cdot \pd \alpha , \nonumber
\\
&-& \frac{f}{2} r \left( \bar{\psi'}^{c}\psi' + \bar{\psi}' {\psi'}^{c}
\right).
\label{eq:fortyfour}
\eeqa
Diagonalization of the $\psi'$ mass matrix  generates the mass eigenvalues,
\begin{equation}
m_\pm =  m_\psi \pm  f  \langle r \rangle, 
\end{equation}
for the two mass eigenstates
\begin{equation}
\psi_- = \frac{i}{\sqrt{2}} \lpa \psi'^c - \psi'  \rpa
  \quad {\rm and}   \quad \psi_+ = \frac{1}{\sqrt{2}}\lpa \psi'^c+\psi' \rpa  \, .
\label{eq:fourtyseven}
\end{equation}
In this basis, the act of charge conjugation on $\psi_\pm$ results in
\beq
\psi^c_\pm =  \psi_\pm.
\label{eq:fourtyeight}
\eeq This tells us that the fields $\psi_\pm$ are Majorana
fermions. The Lagrangian is found to be 
\beqa \mathscr{L}_\psi
&=&\frac{i}{2}\bar{\psi_+}\gamma \cdot \pd \psi_+ +
\frac{i}{2}\bar{\psi_-}\gamma \cdot \pd \psi_- - \frac{1}{2} m_+
\bar{\psi}_+ \psi_+ - \frac{1}{2}m_- \bar{\psi}_- \psi_- , \nonumber
\\
&-&\frac{i}{4 \langle r \rangle} (\bar{\psi}_+ \gamma \psi_- - \bar{\psi}_-
\gamma \psi_+) \cdot \pd \alpha' , \nonumber
\\
& -& \frac{f}{2} r (\bar{\psi}_+\psi_+ - \bar{\psi}_- \psi_-) \,,
\label{eq:fortynine} 
\eeqa where $\alpha' \equiv 2 \alpha \langle r \rangle$ is the canonically normalized
Goldstone boson.  We must now put $r$ into its massive field
representation, for which the interactions of interest are \beq
-\frac{f \sin \theta}{2} H (\bar{\psi}_+\psi_+ - \bar{\psi}_- \psi_-) -
\frac{f \cos \theta}{2} h (\bar{\psi}_+\psi_+ - \bar{\psi}_- \psi_-).
\label{eq:fifty}
\eeq This leads to 3-point interactions between the Majorana fermions and the
Higgs boson of the SM.

In summary, the Dirac fermion of the hidden sector splits into two
Majorana mass-eigenstates. The heavier state will decay into the
lighter one by emitting a Goldstone boson. The lighter one, however,
is kept stable by the unbroken reflection symmetry. Therefore, we can
expect that the universe today will contain only one type of Majorana
WIMP, the lighter one $w$, with mass $m_w$ equal to the smaller of
$m_\pm$. The dark sector hence contains five unknown parameters,
$m_w$, $m_h$, $\lambda$, $\theta$, and $f$. To avoid fine tuning
herein we impose an additional constraint relating some of these free
parameters: $\Delta m/m_w \ll 1$, where \mbox{$\Delta m = |m_+ - m_-| = 2 |f
\langle r \rangle|$.}

\section{Goldstone bosons as imposter fractional neutrinos}
\label{sec3}

In the early Universe, the Goldstone bosons were at thermal
equilibrium with the SM particles. As the Universe cooled due to its
Hubble expansion, $H (T) \simeq 1.66 \sqrt{g(T)} T^2/M_{\rm Pl}$, the
weakly interacting Goldstone bosons decoupled from the SM
particles. Throughout $M_{\rm Pl}$ is the Planck mass and $g(T)$ is
the effective number of interacting (thermally coupled) relativistic
degrees of freedom at temperature $T$. Following Weinberg we require
that the $\alpha$'s go out of equilibrium when the temperature is
about the muon mass, $T_{\alpha'}^{\rm dec} \approx m_\mu \simeq
105~{\rm MeV}$, ensuring that
\begin{equation}
N_{\rm eff} = 3 +  (4/7)(43/57)^{4/3} = 3.39 \, . 
\end{equation}
For $ 0.2~{\rm GeV} \alt m_h \alt 4~{\rm GeV}$, the interaction rate
of Goldstone bosons is
dominated by resonant annihilation into fermion-antifermion pairs~\cite{Garcia-Cely:2013nin}. The $\alpha \alpha \to \bar {\rm f}
{\rm f}$ rate is given by 
\beq
\Gamma (T) =  \frac{g_\theta^2}{256 \ \pi} \ \frac{m_h^6 }{ m_H^4
  \Gamma_h} \
\frac{K_2(m_h/T)}{T^2} \   \sum_{\rm f} m_{\rm f}^2 \left(1- \frac{4
    m_{\rm f}^2}{m_h^2} \right)^{3/2} \ ,
\label{XsX}
\eeq where $K_2 (x)$ is the 2$^{\rm nd}$ Modified Bessel function of
the second kind and $\Gamma_h$ is the decay width of the $CP$-even
scalar; see Appendix~A for details. Note that since the interaction
rate is proportional to the fermion square mass, in (\ref{XsX}) it is
enough to consider only the annihilation into $\mu^\pm$
pairs. Enforcing the decoupling condition, $\Gamma(T^{\rm
  dec}_{\alpha'}) = H(T^{\rm dec}_{\alpha'})$, we obtain \beq
\frac{M_{\rm Pl} \ g_\theta^2}{256 \ \pi} \ \frac{m_h^6}{ m_H^4
  \Gamma_h} \ \frac{K_2(m_h/m_\mu)}{m_\mu^2} \left(1- \frac{4
    m_\mu^2}{m_h^2} \right)^{3/2} \approx 6.28 \, .  \eeq For $m_h < 2
m_w$, the decay width is given by~\cite{Anchordoqui:2013pta} \beq
\Gamma_h = \frac{m_h^3}{32 \, \pi \, \langle r \rangle^2} \approx
\frac{m_h}{16\pi} \, \lambda \ .  \eeq Using (\ref{teta}) we obtain an
expression for the decoupling condition relating the two unknown
parameters \beq \sin \theta \approx \frac{8 \ \sqrt{6.28} \ m_H^2 \
  m_h \ \langle \phi \rangle}{\sqrt{8 \, M_{\rm Pl} \, m_\mu} \ |m_H^2
  - m_h^2| \ m_\mu} \ \left[ \frac{m_h}{m_\mu}
  \left(\frac{m_h^2}{m_\mu^2}- 4 \right)^{3/4} K^{1/2}_2(m_h/m_\mu)
\right]^{-1} \ .  \eeq

\begin{figure}
\begin{center}
\postscript{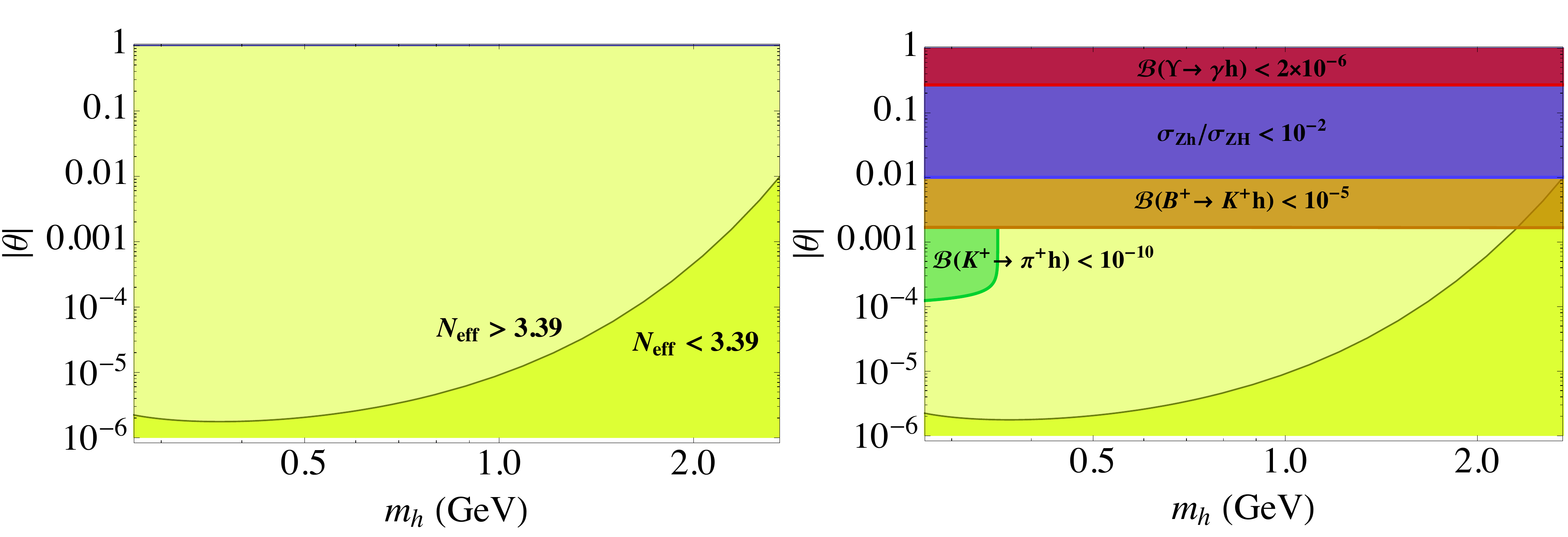}{1.0}
\caption{{\bf Left:} Contour of constant $N_{\rm eff} = 3.39$ in the $(|\theta|,m_h)$ plane.
{\bf Right:} Bounds from interactions involving SM particles in the initial state and the $CP$-even scalar in the final state, overlayed on the same plane.}
\label{fig:w1} 
\end{center}
\end{figure}

 In Fig.~\ref{fig:w1} we show, in the $(|\theta|, m_h)$ plane, the
 contour which corresponds to $N_{\rm eff} = 3.39$.  This particular
 choice of the number of effective neutrino species is midway between
 the value reported by the Planck Collaboration using their best
 determination of $h$ and the value determined using the $h$ observed
 by the Hubble Space Telescope~\cite{Ade:2013lta}. The interesting region lies above the
 contour in the sense that physics beyond the SM would be required.
 In the remainder of the paper we concentrate on constraining this
 region of the parameter space.

\section{Constraints from the hidden scalar}
\label{sec4}

In the spirit of~\cite{Bird:2004ts}, in this section we evaluate the impact of experimental limits on $B^+ \to
K^+ + \met$ reported by the
BaBar~\cite{delAmoSanchez:2010bk,Lees:2013kla,Aubert:2004ws},
CLEO~\cite{Browder:2000qr}, and BELLE~\cite{Lutz:2013ftz}
collaborations, as well as limits on $K^+ \to \pi^+ + \met$ from the
E787~\cite{Adler:2001xv} and E949 experiments~\cite{Anisimovsky:2004hr,Adler:2008zza,Artamonov:2009sz}
on the ($\theta, m_h$) plane.

Before proceeding, we pause to not that combining the upper limit
reproted by the BaBar Collaboration \mbox{${\cal B} (\Upsilon \to
  \gamma + \met) < 2 \times 10^{-6}$~\cite{delAmoSanchez:2010ac}} and
the Wilczek mechanism~\cite{Wilczek:1977zn} with its one loop QCD
correction (which results in $\approx 84\%$ decrease of the total
rate~\cite{Vysotsky:1980cz,Nason:1986tr,Goldberg:1988ji}), one obtains an upper bound for the
mixing angle, $\theta < 0.27$~\cite{Huang:2013oua}. A stronger
constraint follows from LEP limits on the production of
invisibly-decaying Higgs bosons $\sigma_{Zh}/\sigma_{ZH} <
10^{-2}$~\cite{Barate:1999uc,Abdallah:2003ry,Achard:2004cf,Abbiendi:2007ac},
which implies $\theta <
10^{-2}$~\cite{Cheung:2013oya}.

Searches for the rare flavor-changing neutral-current decay $B^+
\rightarrow K^+ + \met$ have been conducted by the
BaBar~\cite{Aubert:2004ws,delAmoSanchez:2010bk,Lees:2013kla},
CLEO~\cite{Browder:2000qr}, and BELLE~\cite{Lutz:2013ftz}
collaborations.  The corresponding SM mode is a decay into $K^+$ and a
pair of neutrinos, with a branching ratio ${\cal B} (B^+ \to K^+ \nu
\bar \nu) \approx 3  \times
10^{-6}$~\cite{Buchalla:2000sk,Bartsch:2009qp}.  No significant excess
of such decays over background has been observed. The most stringent
upper limit  has been reported by the BaBar
Collaboration, ${\cal B} (B^+ \to K^+ +\met) < 1.3 \times
10^{-5}$, at 90\% C.L.~\cite{delAmoSanchez:2010bk}. In our calculation we subtract
the SM contribution to the branching fraction to arrive at 
\begin{equation}
{\cal B} (B^+ \to K^+ h) < 10^{-5}
\label{pulpovidal}
\end{equation}
to be consistent with existing data at 90\% CL.

\begin{figure}[tbp]
\postscript{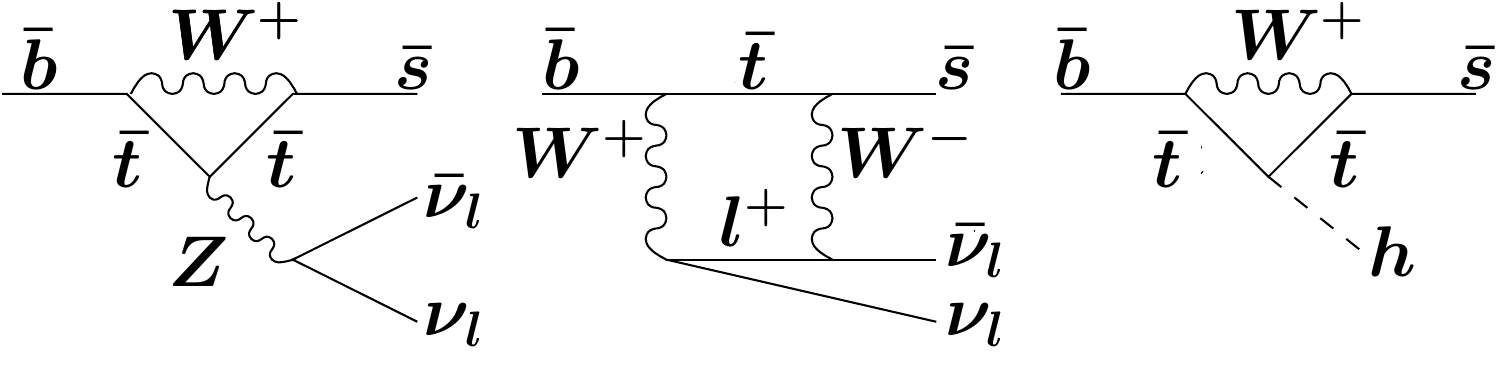}{0.8}
\caption{Feynman diagrams which contribute to $B$ meson decays with
  $\met$. The second-order weak processes that contribute to the $B^+
  \to K^+ \nu \bar \nu$ brancing ratio are the ``$Z$-penguin'' diagram
  (left) and the ``box'' diagram (middle). The hypothetical $b \to sh$
  transition (right) would also yield a missing energy
  signal.}
\label{fig:w2} 
\end{figure}

At the quark level the decays of the $B^+$ meson with missing energy
correspond to the processes shown in Fig.~\ref{fig:w2}. The branching 
fraction for the decay $B^+ \rightarrow K^+ h$ 
is found to be
\begin{eqnarray}
\CB(B^+\rightarrow K^+ h) &=& \frac{9 \sqrt{2} \tau_B G_F^3 m_t^4 m_b^2}{1024 \pi^5 m_B^3 } |V_{tb} V^*_{ts}|^2  \lsb \frac{ (m_B + m_K)(m_B-m_K)}{(m_b - m_s)} f^{BK}_0(m_h^2) \rsb^2 \nonumber \\
&\times& \sqrt{[(m_B + m_K)^2-m_h^2]\, [(m_B - m_K)^2-m_h^2]} \ \theta^2
\ ,
\label{eq:dddy}
\end{eqnarray}
where \beq f^{BK}_0(q^2) = 0.33 \exp \lpa \frac{0.63 q^2}{m_B^2} -
\frac{0.095 q^4}{m_B^4} + \frac{0.591 q^6}{m_B^6} \rpa \eeq is the
form factor~\cite{Abada:2012hf}, $G_F \approx 1.17 \times 10^{-5}~{\rm GeV^{-2}}$ is the Fermi constant, $\tau_B = 1.641 \times
10^{-12}~{\rm s}$ is the $B$-meson lifetime, and $m_B = 5.279~{\rm
  GeV}$, $m_K = 0.494~{\rm GeV}$,  $m_t = 173~{\rm GeV}$, \mbox{$m_b = 4.2~{\rm GeV}$,} and $m_s = 0.095~{\rm GeV}$ are the corresponding particle
masses~\cite{Beringer:1900zz}. The CKM matrix elements yield $|V_{tb} V^*_{ts}| \approx 0.0389$~\cite{Bargiotti:2000dn}. Equating (\ref{pulpovidal}) and
(\ref{eq:dddy}) we obtain an upper limit on the mixing angle \beq
\theta^2 < \frac{1024 \ \pi^5 \ m_B^3}{9 \sqrt{2} \ \tau_B \ G_F^3 \
  m_t^4 \ m_b^2 \ |V_{tb} V^*_{ts}|^2} \frac{10^{-5}}{\sqrt{[(m_B +
    m_K)^2-m_h^2]\, [(m_B - m_K)^2-m_h^2]}} \ \CF_B \, ,
\eeq where \beq \CF_B = \lsb \frac{ (m_B + m_K)(m_B-m_K)}{(m_b - m_s)}
f^{BK}_0(m_h^2) \rsb^{-2} \, .  
\label{roaaar}
\eeq

Lowering the mass of $m_h \alt 355~{\rm MeV}$ opens the decay channel
$K^+ \rightarrow \pi^+ + \met$. Similar to the previous decay process,
the decay $K^+ \rightarrow \pi^+ h$ can proceed through a penguin
diagram.\footnote{Although this penguin have both flippers it appears to be missing both of its legs.} The experiment E949 at Brookhaven National Laboratory studied
the rare decay \mbox{$K^+ \to \pi^+ \nu \bar \nu$} with an exposure of $1.71
\times 10^{12}$ stopped kaons~\cite{Anisimovsky:2004hr,Adler:2008zza,Artamonov:2009sz}. The
data were analyzed using a blind analysis technique yielding five
candidate events.  Combining this result with the observation of two candidate
events by the predecessor experiment E7877~\cite{Adler:2001xv} yields
the branching ratio ${\cal B}(K^+ \to \pi^+ \nu \bar \nu) =
1.73^{+1.15}_{-1.05} \times 10^{-10}$~\cite{Artamonov:2009sz}, which is consistent with the
SM prediction of $(7.81^{+0.80}_{-0.71} \pm 0.29) \times
10^{-11}$ (the uncertainties listed first derive from the input parameters, the smaller uncertainties listed second demonstrate the size of the intrinsic theoretical uncertainties)~\cite{Buchalla:1998ba,Brod:2008ss,Brod:2010hi}. The probability that all seven events were due to background only (background and SM signal) was estimated to be 0.001 (0.073).
In our calculation we subtract the SM branching fraction from the experimental
observation to get 
\begin{equation}
{\cal B} (K^+ \to \pi^+ h) < 10^{-10}
\label{pulpovidalII}
\end{equation}
in order to be consistent with existing data. 
 
Using (\ref{eq:dddy}) with
the appropriate replacements for the quark constituents we obtain the
branching for $K^+ \rightarrow \pi^+ h$ \beqa
\CB(K^+\rightarrow \pi^+ h) &=& \frac{9 \sqrt{2} \tau_K G_F^3 m_t^4 m_s^2}{1024 \pi^5 m_K^3} |V_{ts} V^*_{td}|^2 \lsb \frac{(m_K + m_\pi)(m_K-m_\pi)}{m_s-m_d}f^{K\pi}_0(m_h^2) \rsb^2 \nonumber \\
&\times& \sqrt{[(m_K + m_\pi)^2-m_h^2]\, [(m_K - m_\pi)^2-m_h^2]} \ 
\theta^2 \ ,
\label{eq:popo}
\eeqa
where the form factor is given by~\cite{Ghorbani:2013yh} 
\beq
f_0^{K\pi}(q^2) \approx 0.96 \lpa 1+ 0.02 \frac{q^2}{m_\pi^2} \rpa \, .
\eeq
Here, $\tau_K = 1.24 \times 10^{-8} \
{\rm s}$, $m_\pi = 0.1396~{\rm GeV},$ $m_d = 0.0048~{\rm
  GeV}$, and $|V_{ts} V^*_{td}| = 3.07 \times 10^{-4}$.
Equating (\ref{pulpovidalII}) to (\ref{eq:popo}) we obtain an upper limit on the mixing angle
\beq
\theta^2 <  \frac{1024 \ \pi^5 \ m_K^3}{9 \sqrt{2} \ \tau_K \ G_F^3 \
  m_t^4 \ m_s^2  \ |V_{ts} V^*_{td}|^2} \frac{10^{-10}}{\sqrt{[(m_K +
    m_\pi)^2-m_h^2] \, [(m_K - m_\pi)^2-m_h^2]}} \ \CF_K \, 
\eeq
where
\beq
\CF_K = \lsb \frac{ (m_K + m_\pi)(m_K-m_\pi)}{(m_s - m_d)}
f^{K\pi}_0(m_h^2) \rsb^{-2} \ .
\eeq

In Fig.~\ref{fig:w1} we summarize these results in the ($\theta, m_h$)
plane. Note that for $m_h < 355$~MeV the $K^+ \rightarrow \pi^+ h$
channel dominates, requiring $\left| \theta \right| < 10^{-4}$.  For
355~MeV~$< m_h <$~2 GeV, the $B^+ \rightarrow K^+ h$ dominates,
setting an upper bound $\left| \theta \right| < 10^{-3}$.

It is important to explain the reason the bound derived from $B$
  decay measurements shown in Fig.~\ref{fig:w1} appears not to depend
  on $m_h$.  Recall we are probing regions for which $0.2 {\rm GeV}
  \alt m_h \alt 2~{\rm GeV}$. For these values, the exponential in the
  form factor and the demoninator with the square root, see
  Eq.~(\ref{roaaar}), do not vary much with $m_h$, since they do not
  depend directly on $m_h$ but rather on the ratio $m_h/m_B$ and
  $m_h/(m_B \pm m_K)$, respectively. Since $m_B \sim 5.3~{\rm GeV}$,
  these ratios are small and there is little variation of the bound
  with $m_h$.

\section{Constraints from  the hidden fermions}
\label{sec5}

\begin{figure}[h]
\begin{center}
\postscript{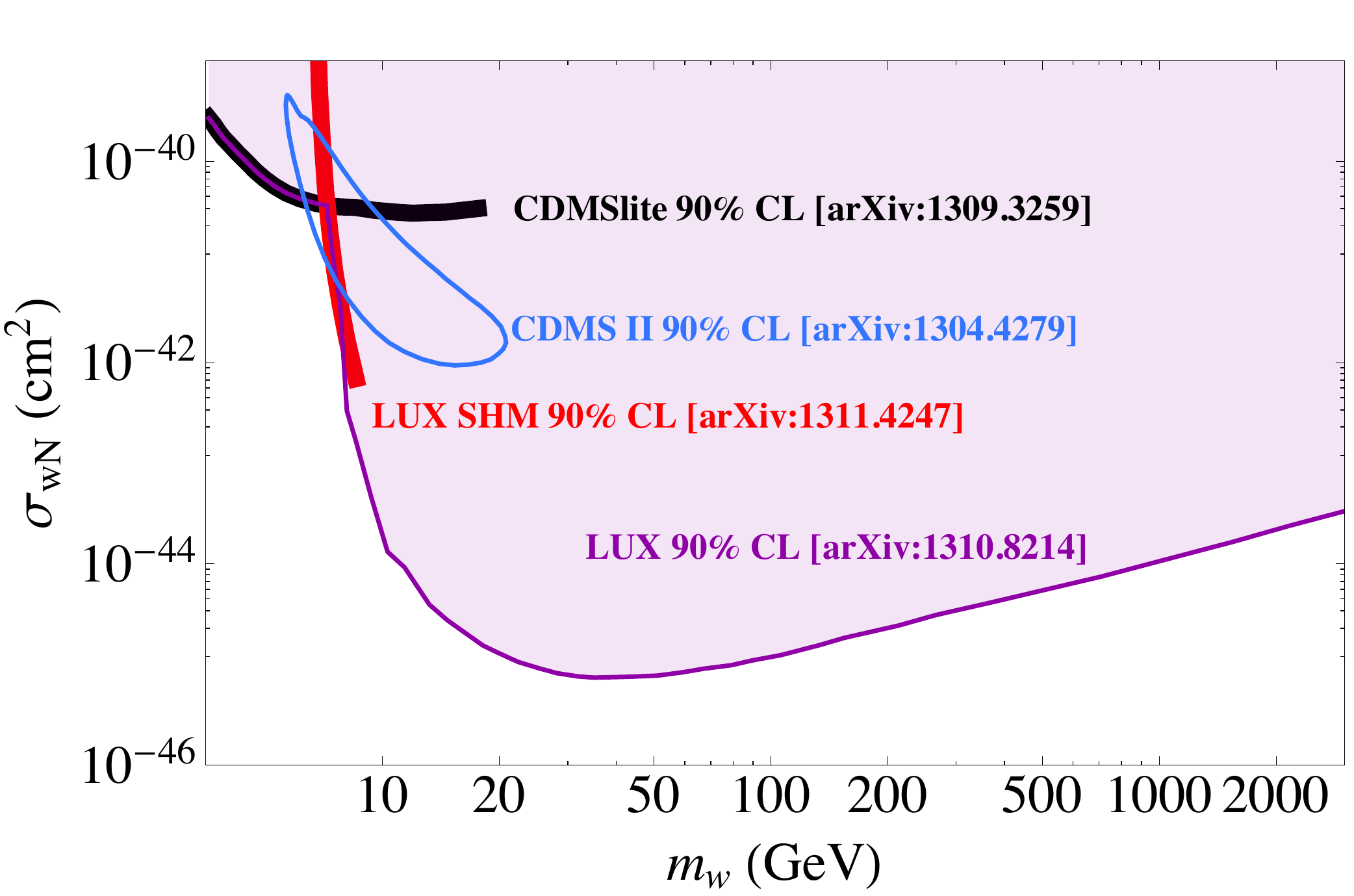}{0.8}
\caption{Favored 90\% C.L. region of CDMS II~\cite{Agnese:2013rvf}, and 90\% C.L. exclusion contours of LUX~\cite{Akerib:2013tjd}, CDMSlite~\cite{Agnese:2013lua}, and LUX assuming the standard halo model (SHM)~\cite{DelNobile:2013gba} in the $(\sigma_{wN},m_w)$ plane.}
\label{fig:w3} 
\end{center}
\end{figure}

Next, in line with our stated plan, we use data from DM searches at
direct detection experiments to constrain the parameter space of the
fermion sector. The WIMP-nucleon cross section for scalar interactions
is found to be~\cite{Beltran:2008xg}.  \beq \sigma_{wN} =
\frac{4}{\pi} \frac{m_w^2 m_N^4}{(m_w + m_N)^2} f_N^2 \ , \eeq where
$m_N \simeq 1~{\rm GeV}$ is the nucleon mass and $f_N$ is the
WIMP-nucleon coupling.  For the case at hand,
\begin{equation}
f_N \simeq  \frac{G_q}{4 m_q}, \quad \quad  {\rm with} \quad \quad  \frac{G_q}{m_q} = \frac{2 f g_\theta
\langle r \rangle}{2 \sqrt{2} m_H^2 m_h^2}, 
\end{equation}
yielding~\cite{Anchordoqui:2013pta} 
\beq
\sigma_{wN} = \left(\frac{1}{2\sqrt{2}} \right)^{2} \frac{1}{4\pi}
\frac{m_w^2 m_N^4}{(m_w + m_N)^2} \left( \frac{2 g_\theta \, \langle r
    \rangle \, f}{ m_H^2 m_h^2} \right)^2 \ .
\eeq
We may re-express this result in terms of the mixing angle,
\beq
\sigma_{wN} = (0.35)^{2}  \frac{1}{4\pi}  \frac{m_w^2 m_N^4}{(m_w +
  m_N)^2} \left( \frac{ f}{ \langle \phi \rangle} \right)^2 \left( \frac{1}{m_H^2} - \frac{1}{m_h^2} \right)^2 \sin^2 2 \theta \ .
\eeq
For $\theta \ll 1$, the upper limits on the nucleon-wimp cross
sections derived by the various experiments translate into upper
limits of the mixing angle
\beq
|\theta| < \frac{(m_w + m_N)}{m_N^2 m_w} \frac{\langle \phi \rangle}{f}  \left| \frac{1}{m_H^2} - \frac{1}{m_h^2} \right|^{-1} \frac{\sqrt{\pi}}{0.35} \sqrt{ \sigma_{wN}(m_w) }  \ .
\label{DM_bound}
\eeq

\begin{figure}[tbp]
\begin{minipage}[t]{0.49\textwidth}
\postscript{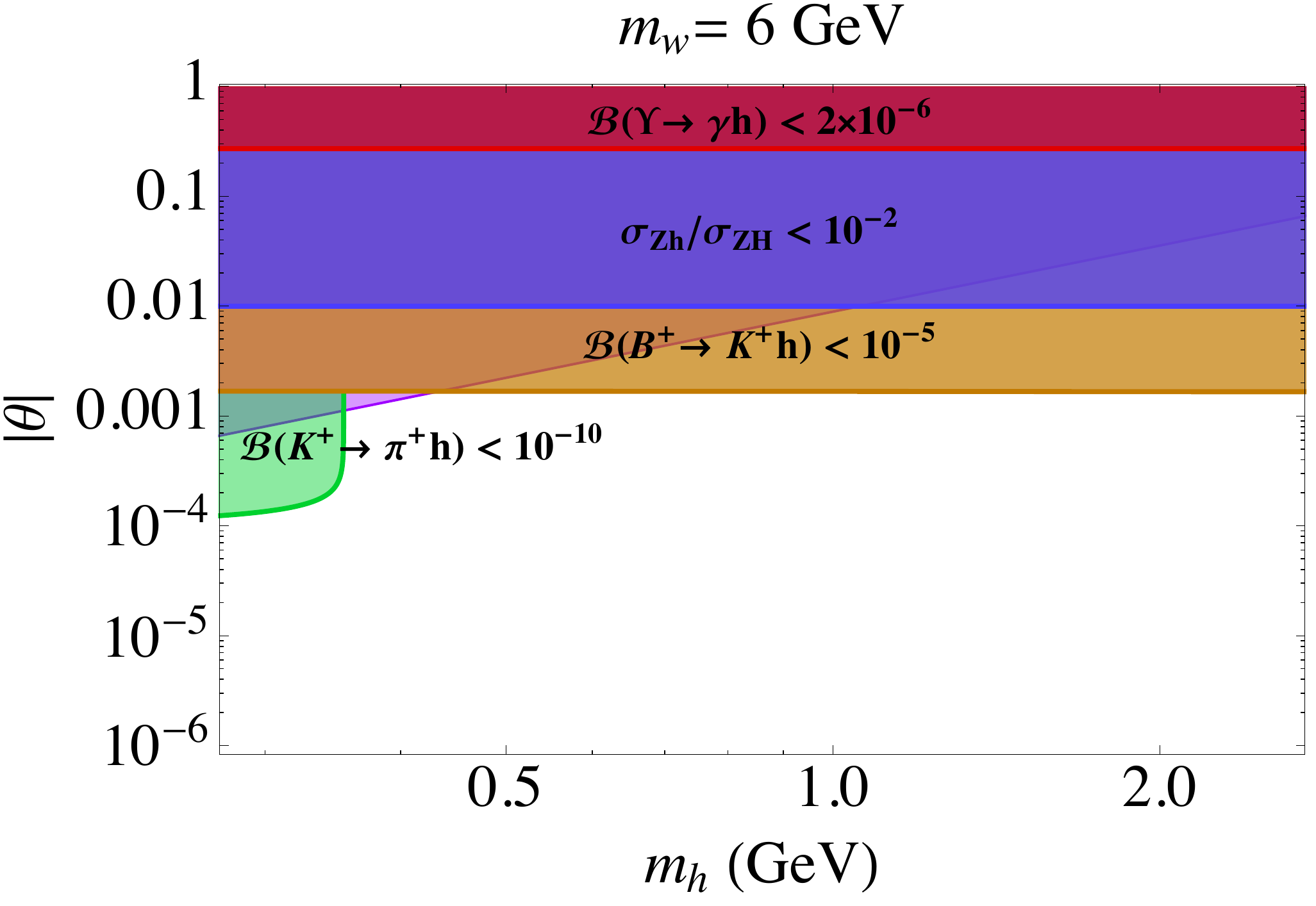}{0.99}
\end{minipage}
\hfill
\begin{minipage}[t]{0.49\textwidth}
\postscript{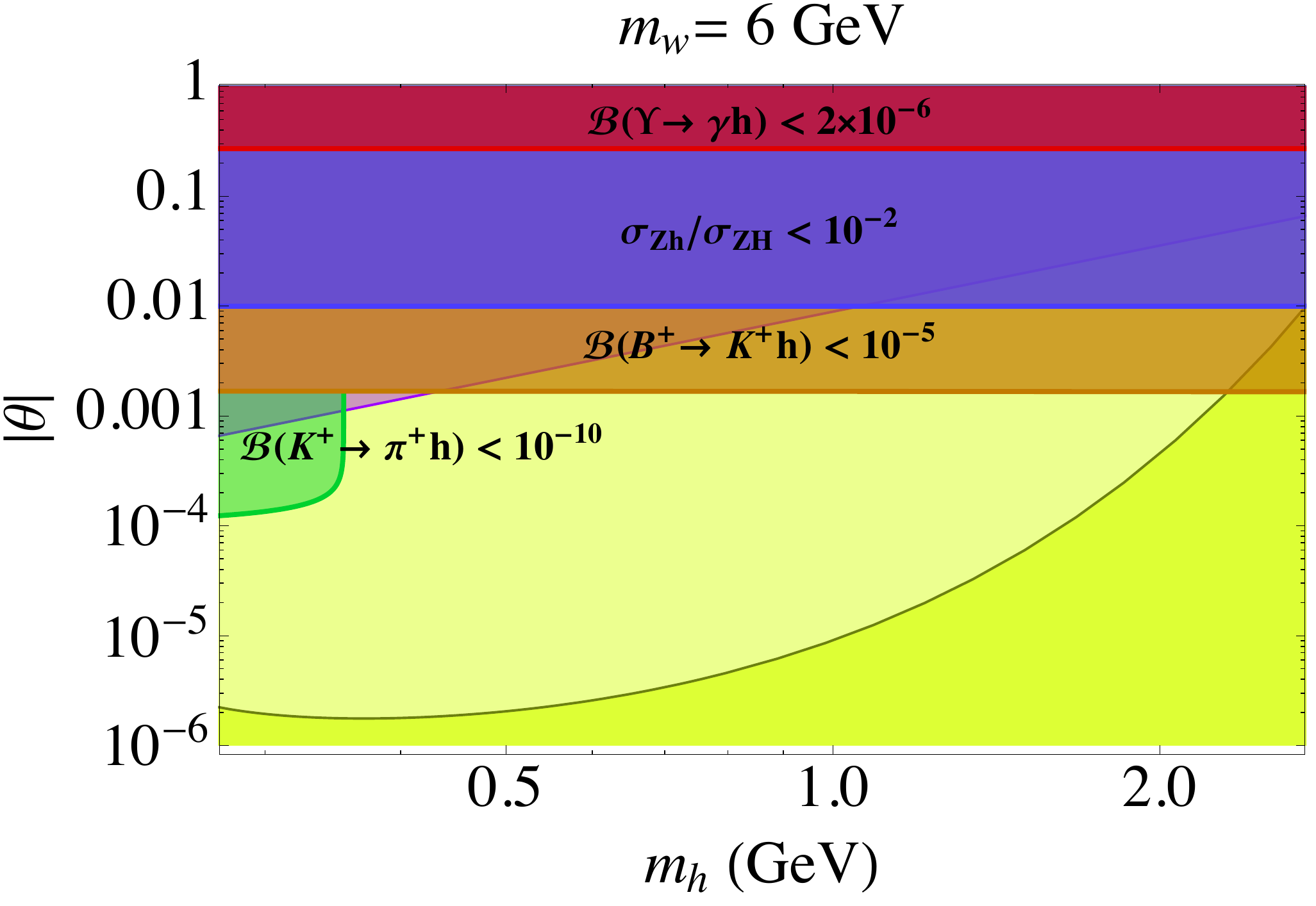}{0.99}
\end{minipage}
\begin{minipage}[t]{0.49\textwidth}
\postscript{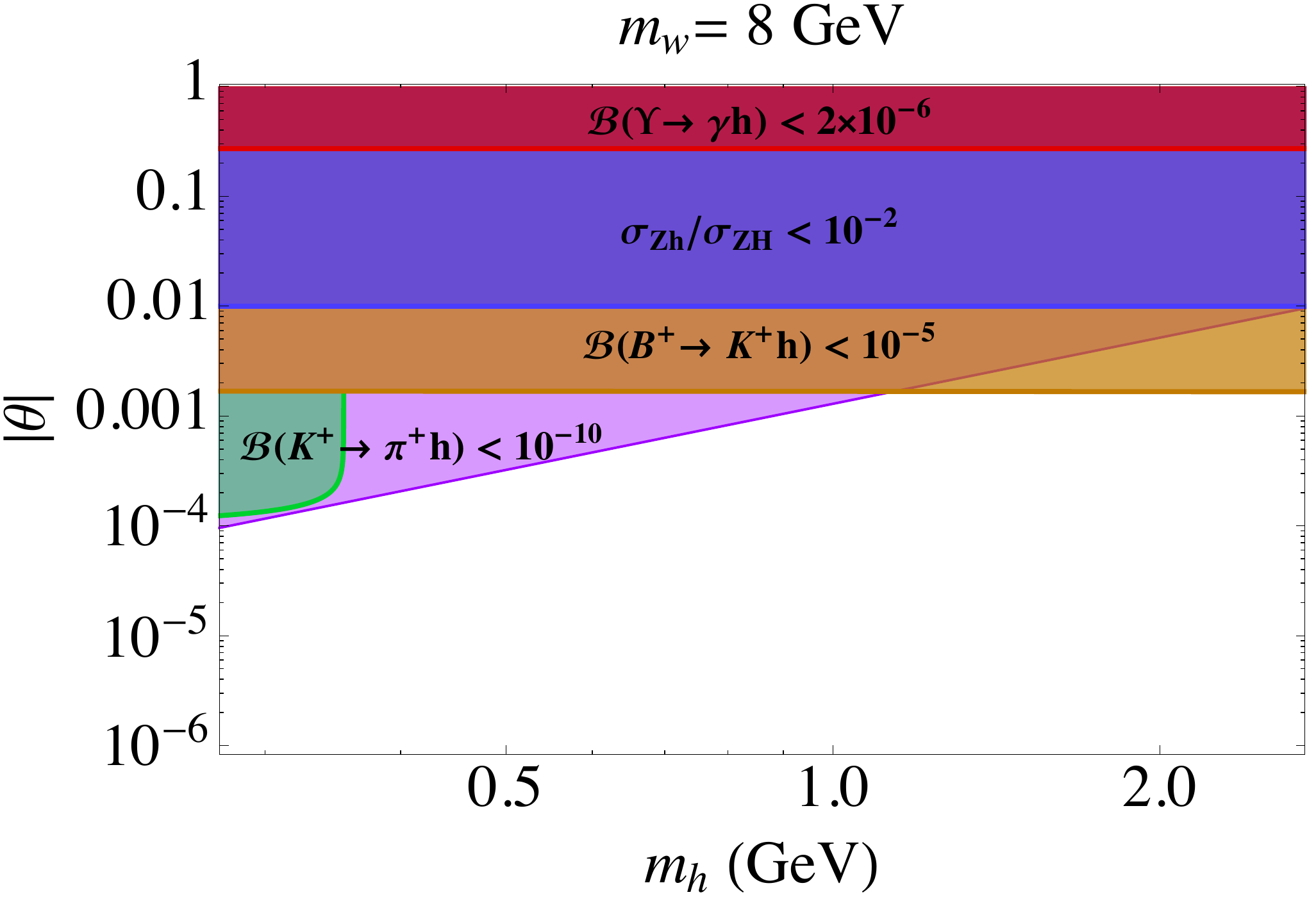}{0.99}
\end{minipage}
\hfill
\begin{minipage}[t]{0.49\textwidth}
\postscript{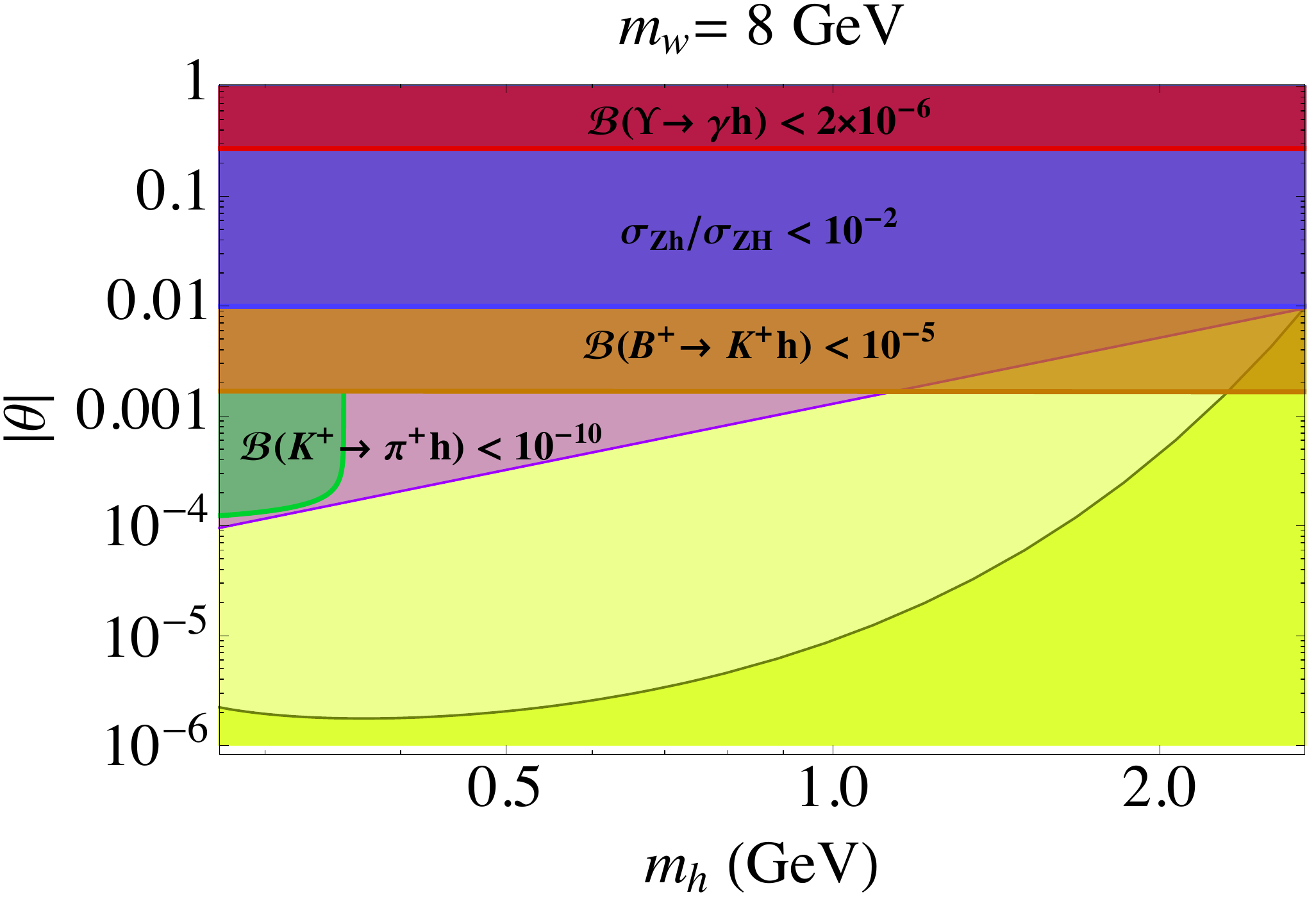}{0.99}
\end{minipage}
\begin{minipage}[t]{0.49\textwidth}
\postscript{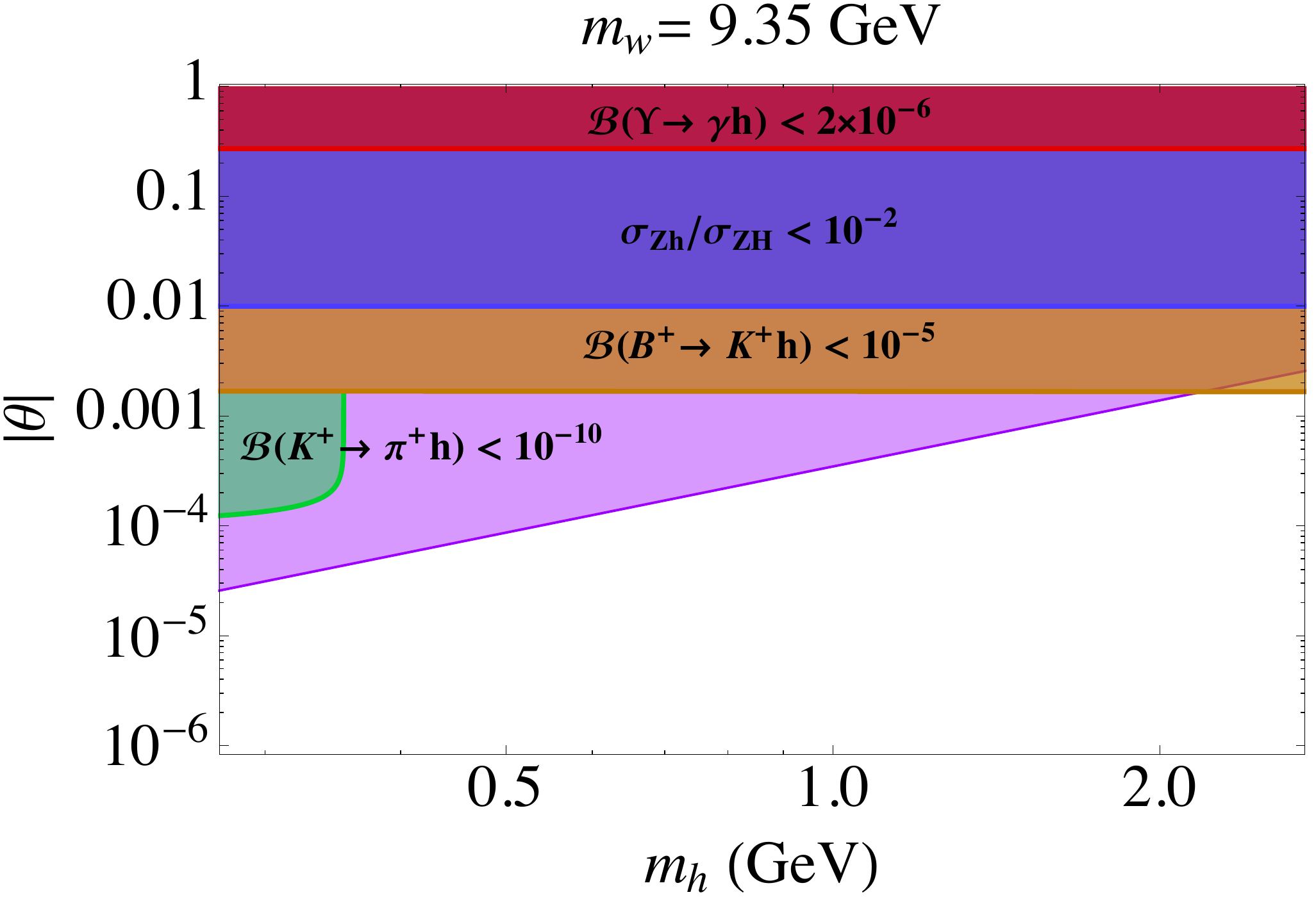}{0.99}
\end{minipage}
\hfill
\begin{minipage}[t]{0.49\textwidth}
\postscript{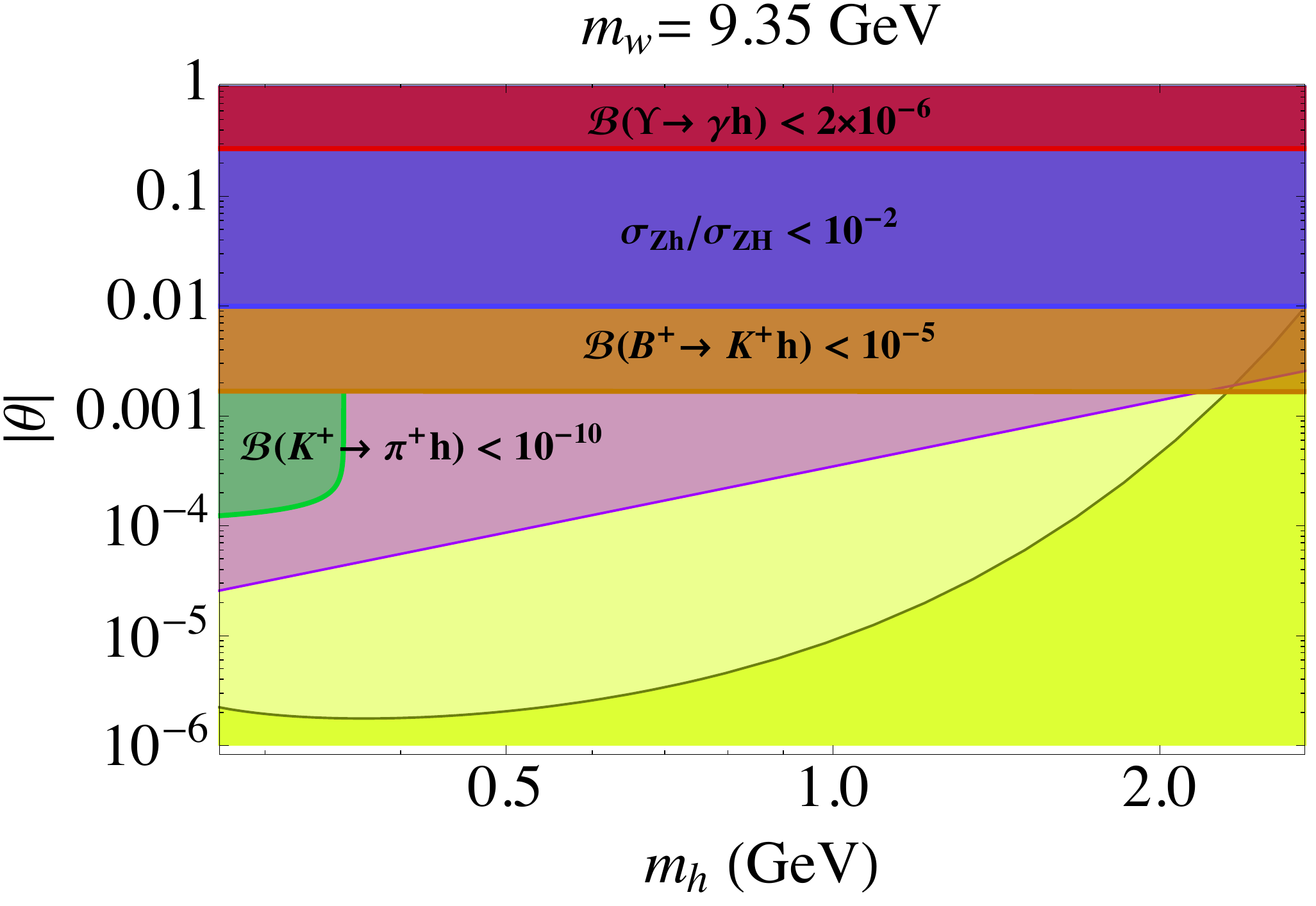}{0.99}
\end{minipage}
\caption{Plots on the left indicate bounds in the ($\theta, m_h$) plan arising 
from heavy meson decays with missing energy as well as bounds from DM direct detection
experiments.  On the right, the contour for $N_{\rm eff} = 3.39$ is overlayed on the bounds. We have taken $m_w = 6~{\rm GeV}$, 8~GeV, and 9.35~GeV.}
\label{fig:w4} 
\end{figure}

\begin{figure}[tbp]
\begin{minipage}[t]{0.49\textwidth}
\postscript{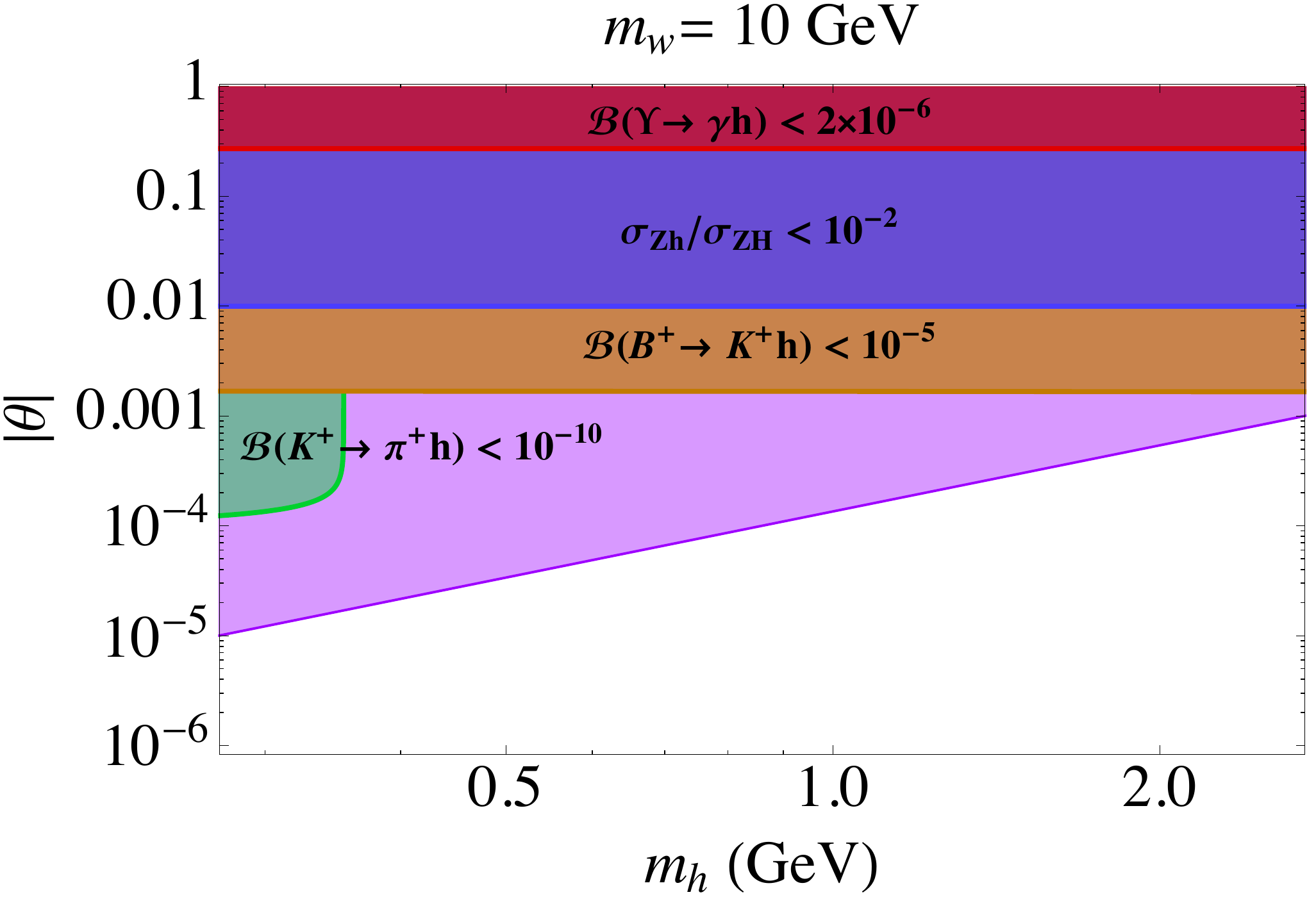}{0.99}
\end{minipage}
\hfill
\begin{minipage}[t]{0.49\textwidth}
\postscript{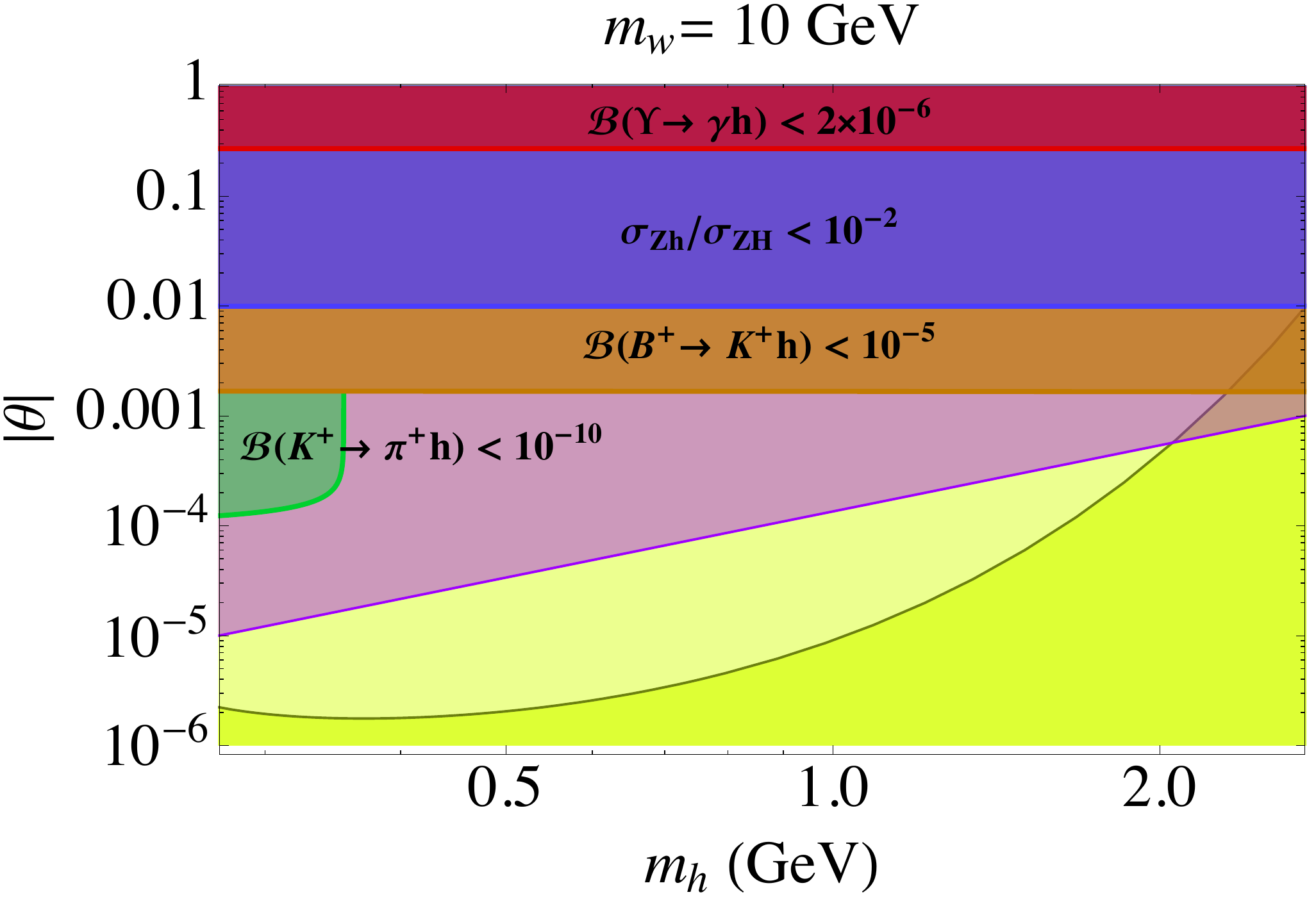}{0.99}
\end{minipage}
\begin{minipage}[t]{0.49\textwidth}
\postscript{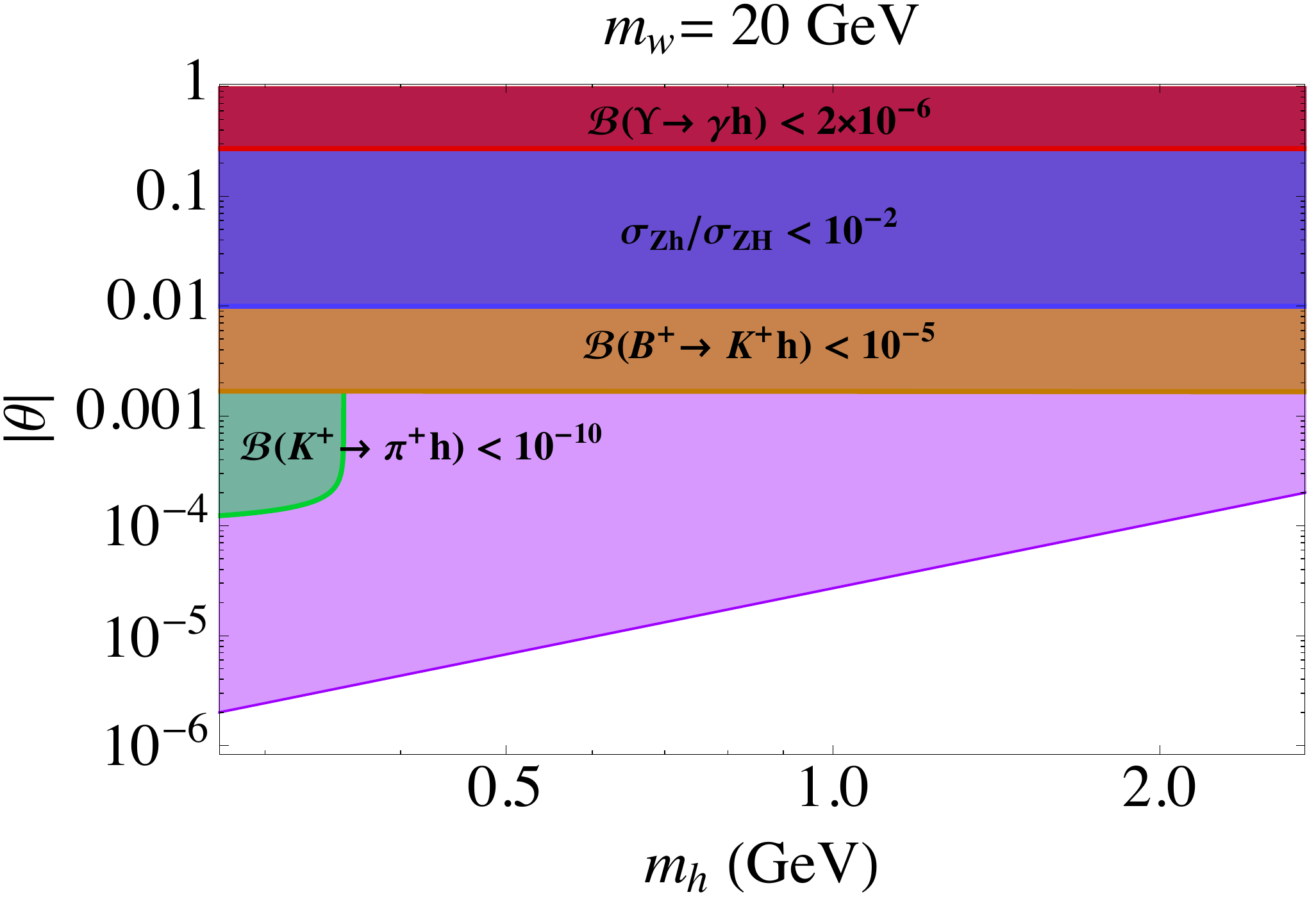}{0.99}
\end{minipage}
\hfill
\begin{minipage}[t]{0.49\textwidth}
\postscript{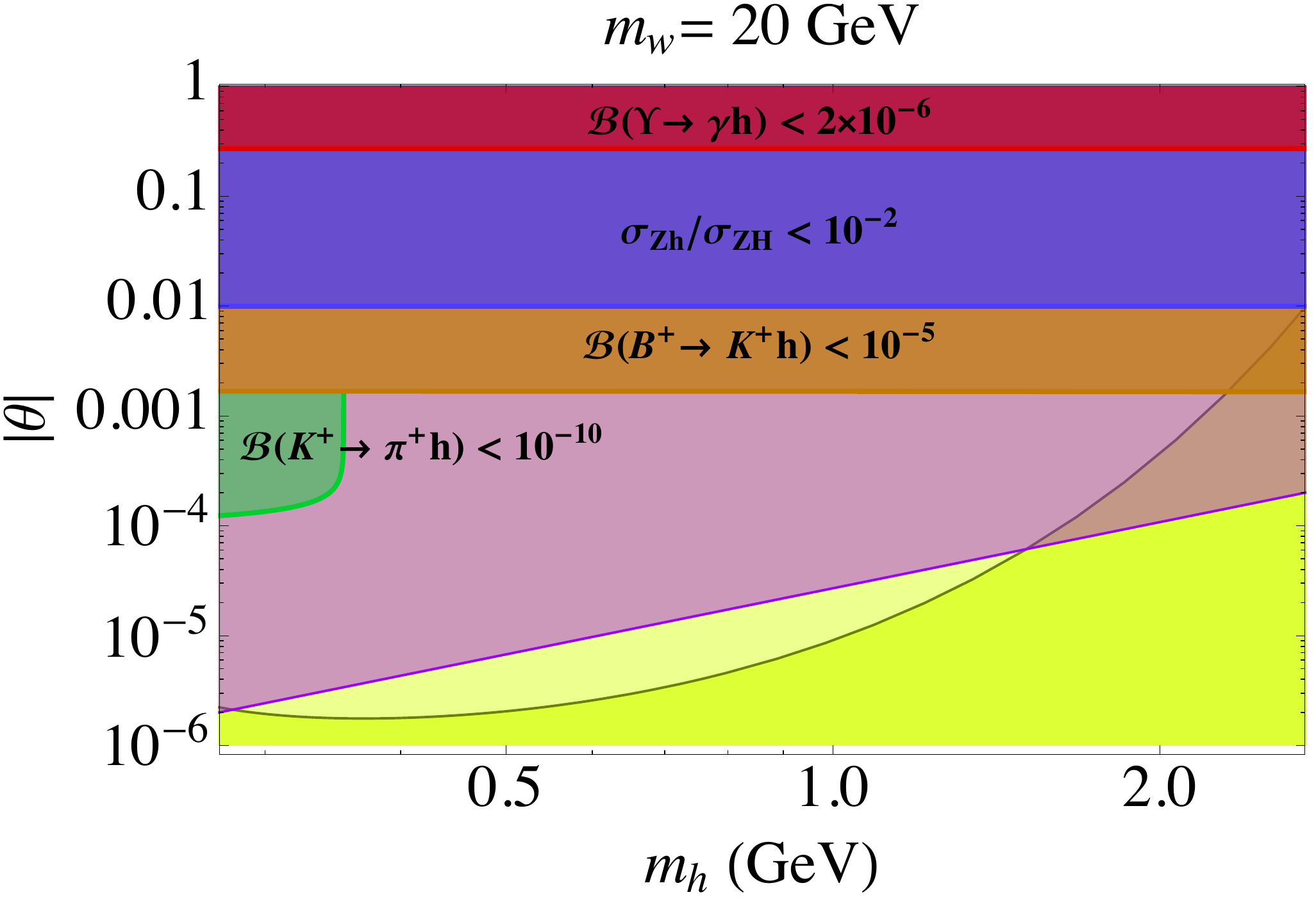}{0.99}
\end{minipage}
\begin{minipage}[t]{0.49\textwidth}
\postscript{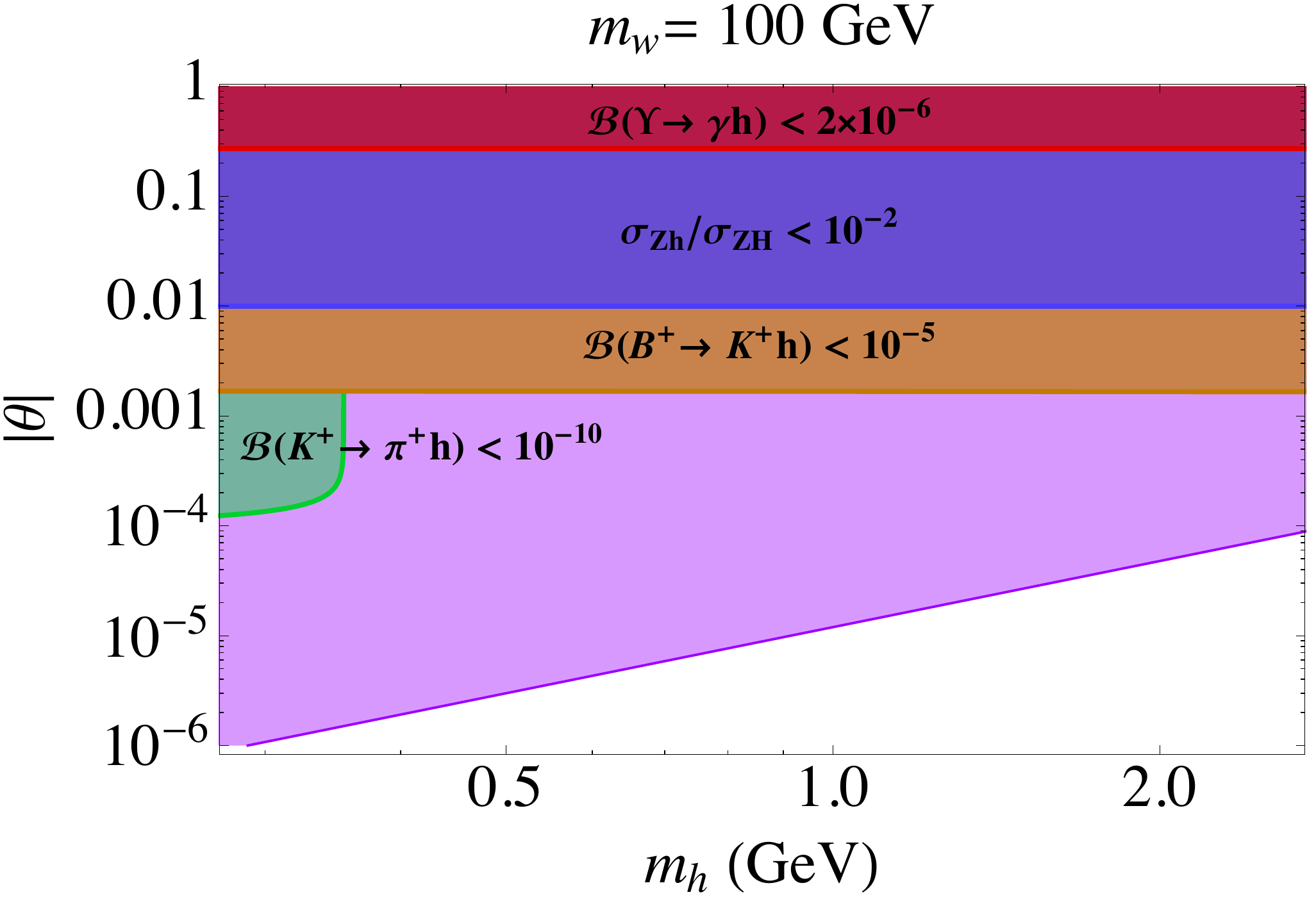}{0.99}
\end{minipage}
\hfill
\begin{minipage}[t]{0.49\textwidth}
\postscript{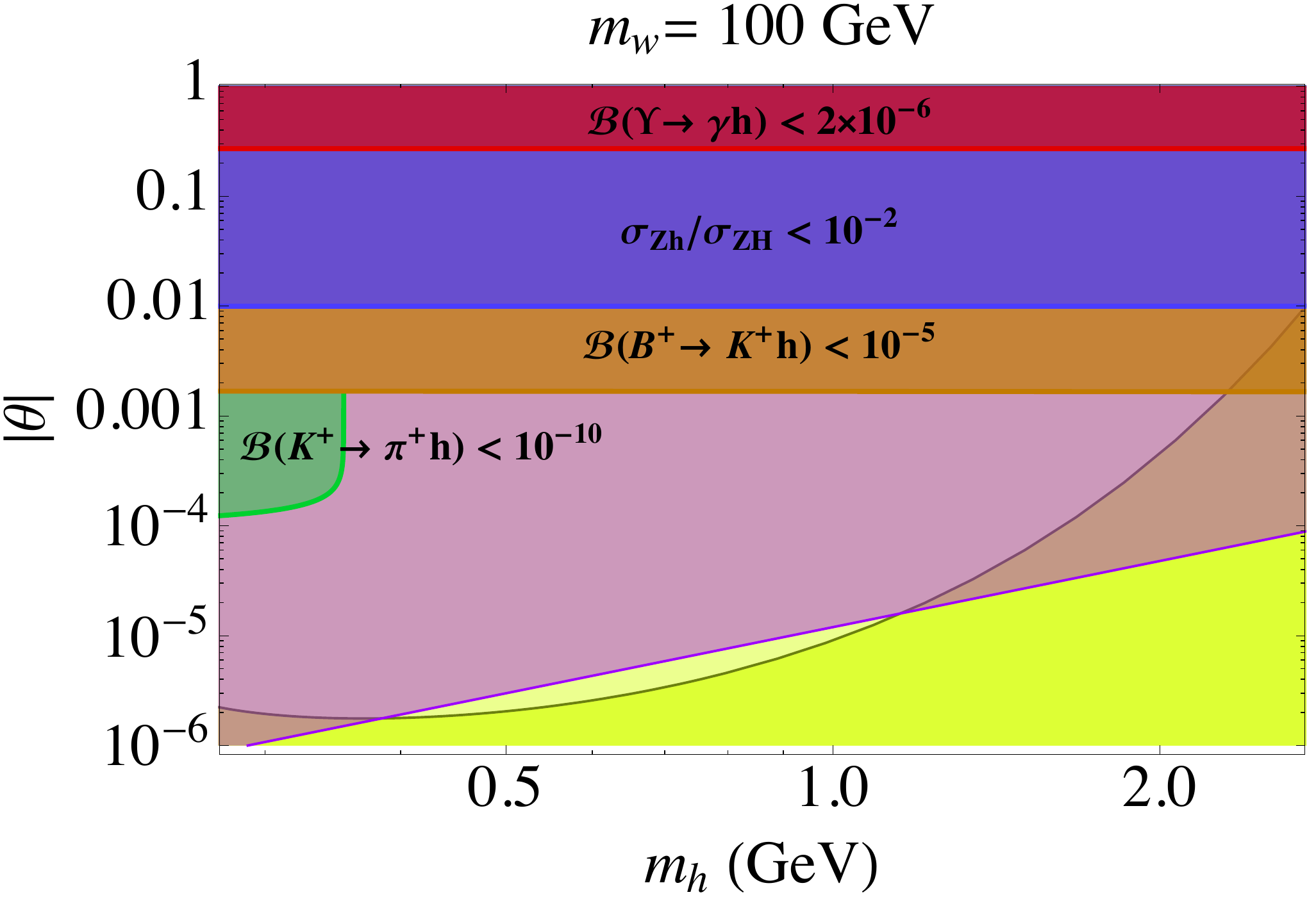}{0.99}
\end{minipage}
\begin{minipage}[t]{0.49\textwidth}
\postscript{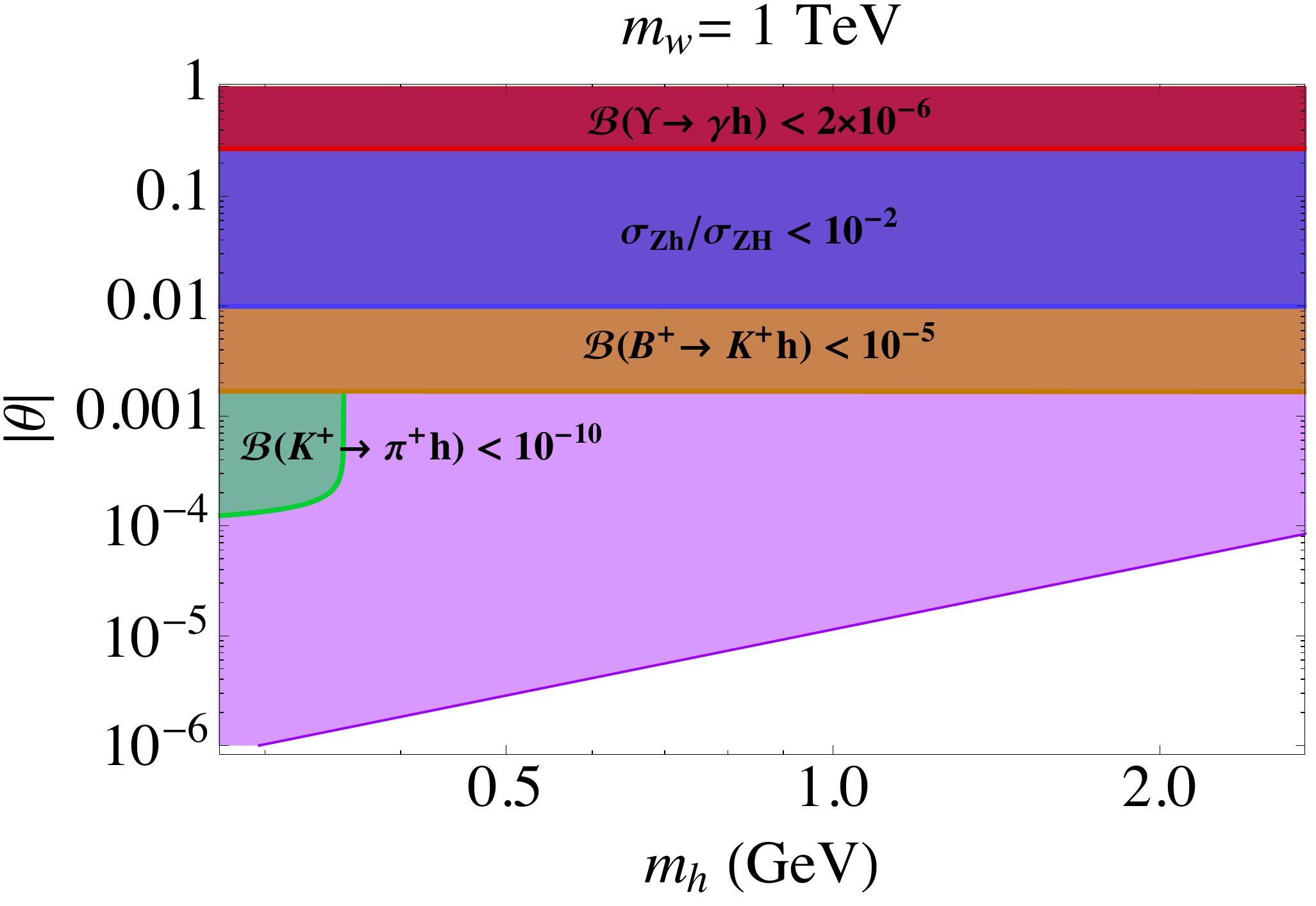}{0.99}
\end{minipage}
\hfill
\begin{minipage}[t]{0.49\textwidth}
\postscript{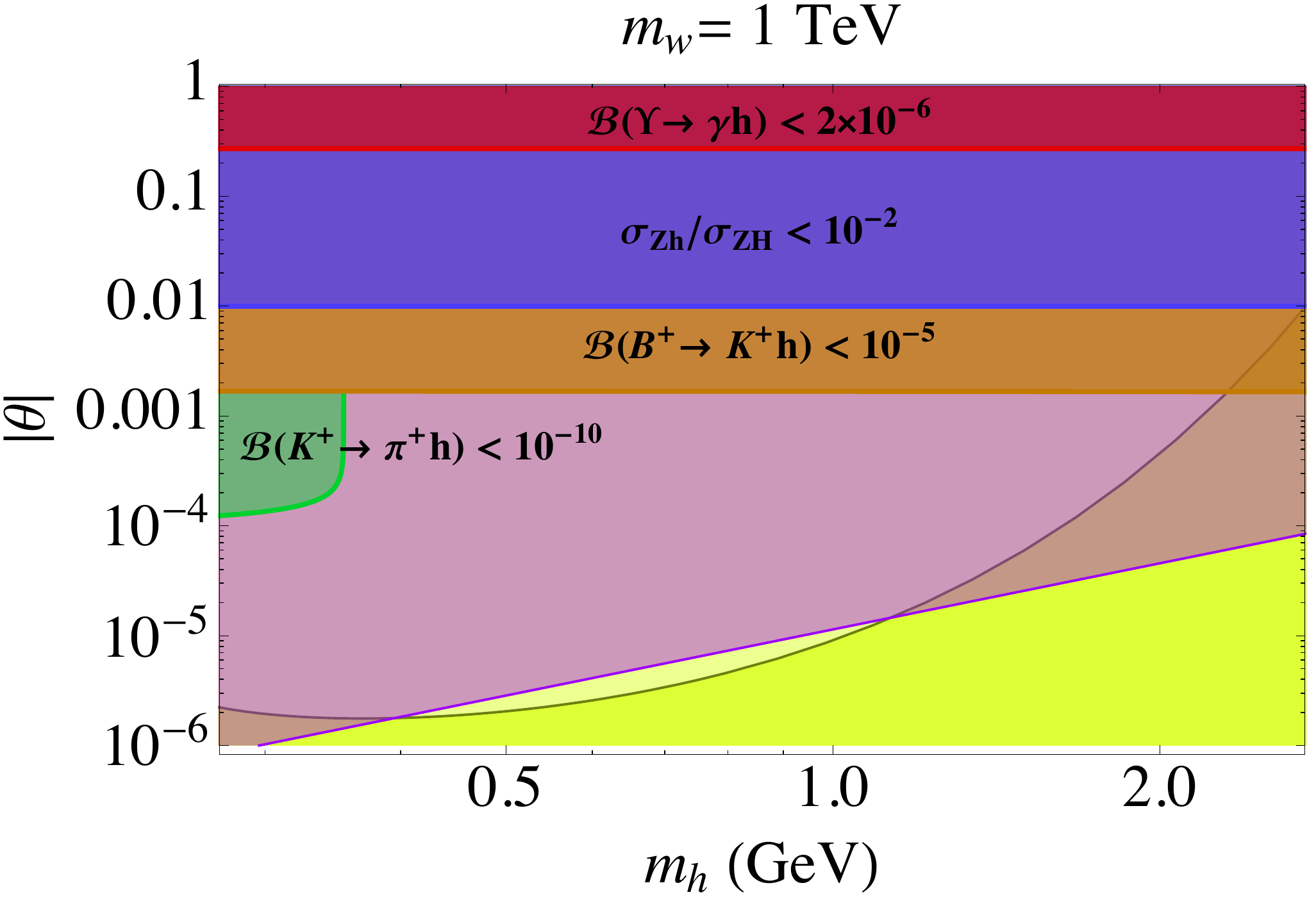}{0.99}
\end{minipage}
\caption{Idem Fig.~\ref{fig:w4}, but for different values of $m_w$.}
\label{fig:w5} 
\end{figure}

To determine $f$ we require the $w$ relic density to be consistent
with $h^2 \Omega_{\rm DM} \simeq 0.111(6)$.  In our study we consider
the interesting case in which $m_h < m_w$, for which threshold andf resonant effects are negligible and thus the
instantaneous freeze-out approximation is valid~\cite{Garcia-Cely:2013nin}.  In this region of
the parameter space, the $w$'s predominantly annihalate into a pair of $h$'s or co-annihilate with the next-to-lightest
Majorana fermion, producing a scalar $h$ and a Goldstone boson.  All
of the final state $h$'s subsequently decay into $\alpha$'s.  We note,
however, that for $m_w \approx m_H/2$ one expects dominant
annihilation into fermions. Indeed, resonant annihilation of $w$ into
fermion and subsequent photon production has been proposed as a
possible DM signal accesible to $\gamma$-ray
detectors~\cite{Anchordoqui:2013pta}.\footnote{An alternative
  $\gamma$-ray signal has been proposed in~\cite{Baek:2013ywa}.}
Interestingly, for $m_w \simeq 60~{\rm GeV}$, resonant Higgs
production will result in predominantly $b \bar b$ final states, which
in turn hadronize to states including photons that may be consistent
with the photon flux in the Fermi
bubbles~\cite{Hooper:2013rwa,Huang:2013pda,Okada:2013bna,Huang:2013apa}.

We compute the  thermal-angular average using the 
Gondolo-Gelmini technique~\cite{Gondolo:1990dk}, \beqa \la \sigma_{ww}
v_M \ra &=& \frac{1}{n_w^2(T)} \int \frac{d^3 p_1}{(2\pi)^3} \frac{d^3
  p_2}{(2\pi)^3} \sigma_{ww} v_M e^{-E_1/T}
e^{-E_2/T} \ , \nonumber \\
&=& \frac{x_f}{8 m_w^5 K_2^2(x_f)} \int_{4 m_w^2}^\infty \ \sigma_{ww} (s)
\sqrt{s}(s-4m_w^2) K_1(x_f \sqrt{s}/m_w ) \ ds \,, \eeqa where $x_f =
m_w/T_f$, with $T_f$ the freeze-out temperature.  It is easily seen
that~\cite{Garcia-Cely:2013nin} \beq \lim_{\Delta m /m_w
  \rightarrow 0} \la \sigma_{ww} v_M \ra \approx \frac{f^4}{32 \pi
  m_w^2} \ .
\label{notsoeasy}
\eeq
The freeze out analysis of the Boltzmann equation gives,
\beq
\la \sigma_{ww}  v_M(x_f) \ra=   \frac{1.04 \times 10^9  \, x_f}{\sqrt{g(x_f)} \,  M_{\rm Pl} \Omega_{\rm DM} h^2}~{\rm GeV}^{-1}  \ ,
\label{eq:Dec3}
\eeq
which, for pedagogical reasons, is derived in detail in Appendix~B.
Combining (\ref{notsoeasy}) and (\ref{eq:Dec3}) we obtain
\begin{equation}
\frac{f^4}{32 \pi m_w^2} =  \frac{1.04 \times 10^9~{\rm GeV^{-1}} \, x_f}{\sqrt{g(x_f)} \  M_{\rm Pl} \, \Omega_{\rm DM} h^2}  \ , 
\end{equation}
or equivalently
\begin{equation}
f \approx \left( \frac{1.04 \times 10^{11}~{\rm GeV^{-1}} \, x_f}{\sqrt{g(x_f)} \ M_{\rm Pl} \, \Omega_{\rm DM} h^2} \right)^{1/4} \sqrt{m_w}  \ .
\end{equation}

Since the WIMPs couple to the SM via the Higgs, the model is
isospin-invariant.\footnote{For a related isospin violating dark matter
  model see~\cite{Okada:2013cba}.}  Thus, to determine bounds on the $(|\theta|,m_h)$
parameter space, we consult the experimental limits on the WIMP
nucleon cross section shown in Fig.~\ref{fig:w3}.  Placing these limits in
(\ref{DM_bound}) together with the value of $f$ derived from the
requirement that the relic density is correctly reproduced, we extract
limits on $|\theta |$, as shown in Figs.~\ref{fig:w4} and
\ref{fig:w5}. Once $m_w$ exceeds 8~GeV, the bounds from direct
detection experiments begin to constrain the parameter space.  For
$m_w > 9.35~{\rm GeV}$ the bounds from direct detection dominate over the bounds
from the interactions involving the $CP$-even scalar.  Recall that this
analysis does not account for excitations of the SM Higgs, which
prevents us from using this technique to probe regions where $55~{\rm GeV} \alt m_w
\alt 70~{\rm GeV}$.  For $m_w > 100~{\rm GeV}$, the region requiring new physics has
been nearly excluded.

\section{Constraints from Higgs decay into invisibles}
\label{sec6}

\begin{figure}[h]
\begin{center}
\postscript{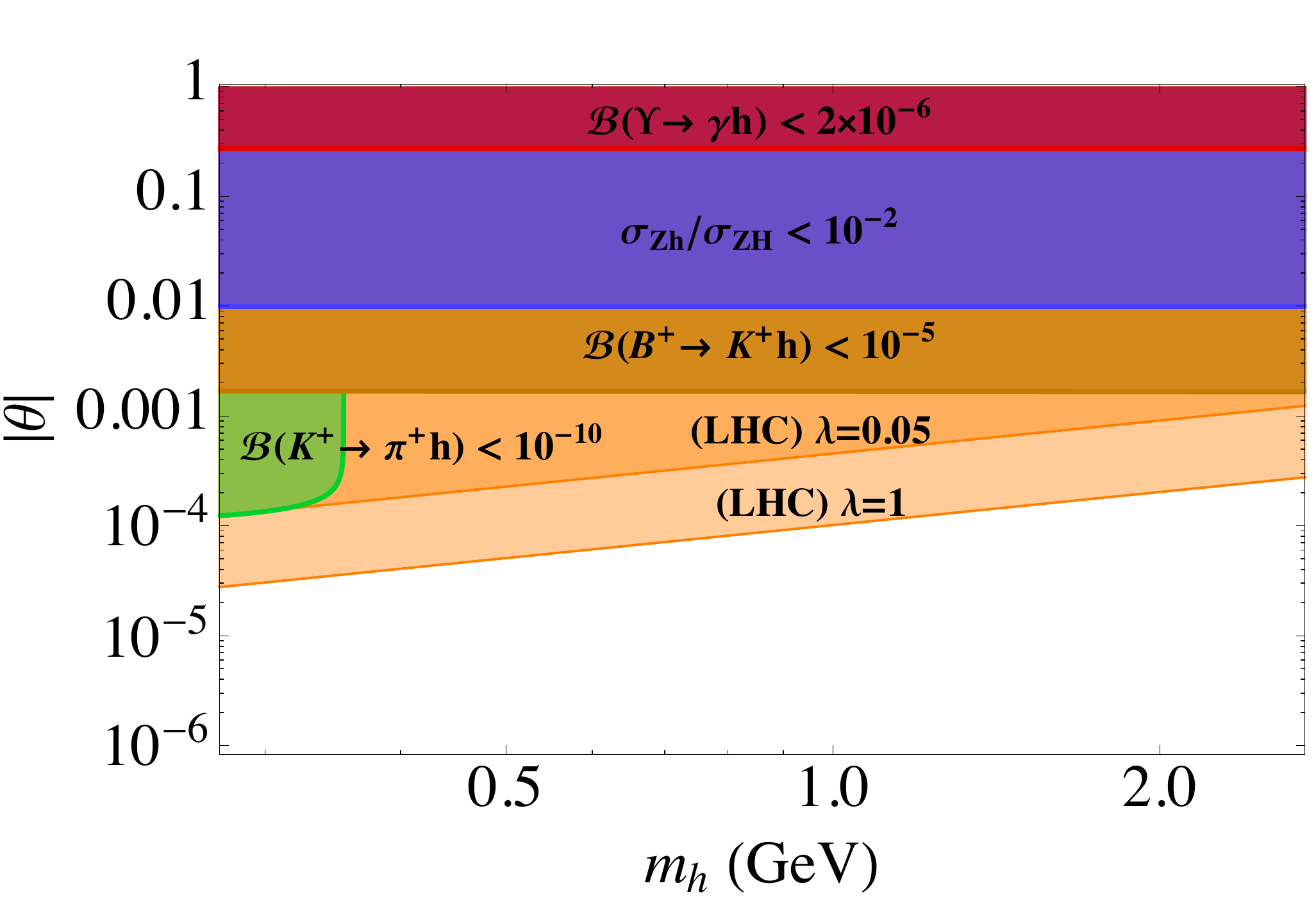}{0.7}
\caption{Bounds on the $(\theta, m_h)$ including invisible Higgs decays for different assumptions about the value of the quartic 
coupling $\lambda$.}
\label{fig:w6} 
\end{center}
\end{figure}

\begin{figure}[tbp]
\begin{minipage}[t]{0.49\textwidth}
\postscript{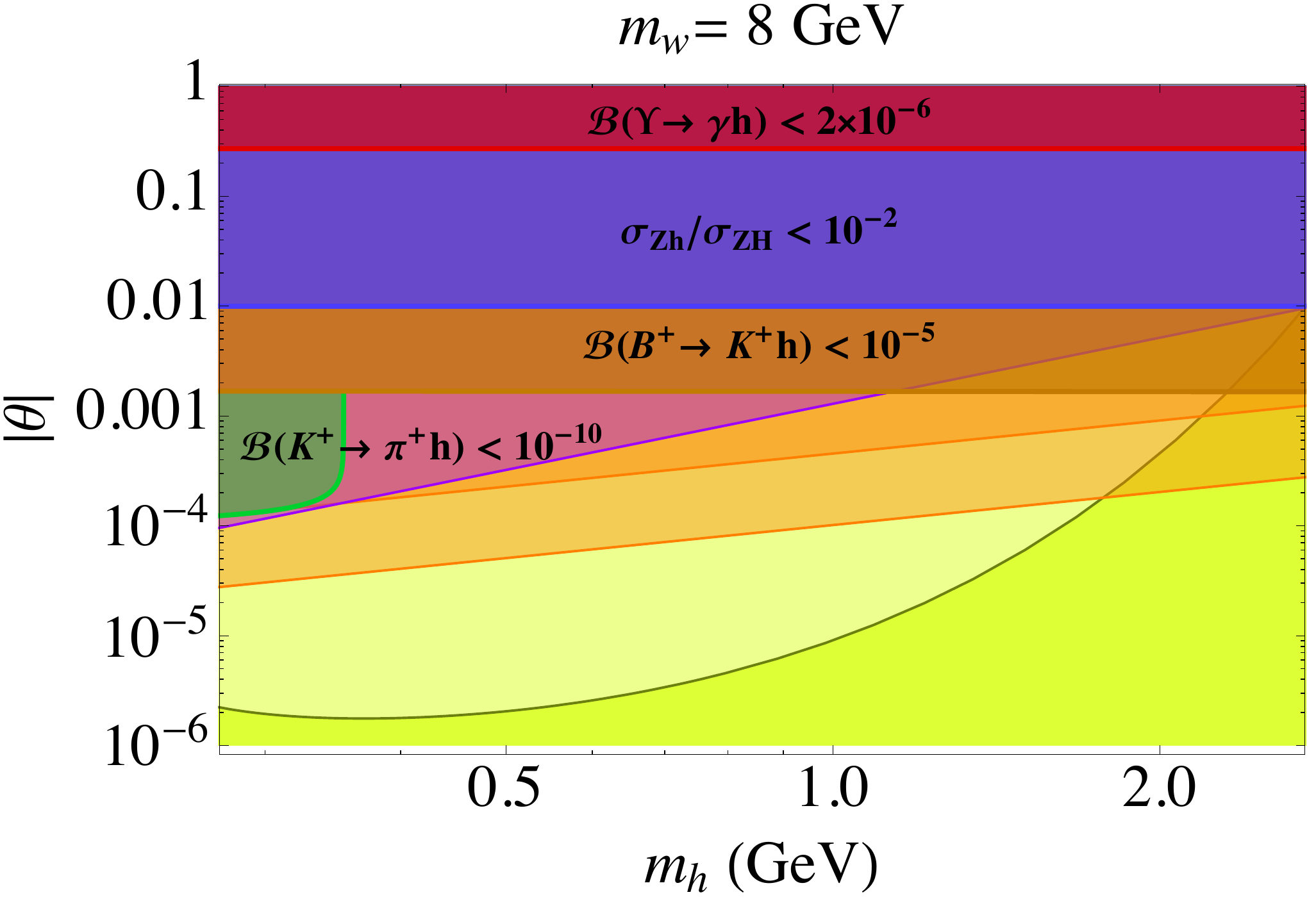}{0.99}
\end{minipage}
\hfill
\begin{minipage}[t]{0.49\textwidth}
\postscript{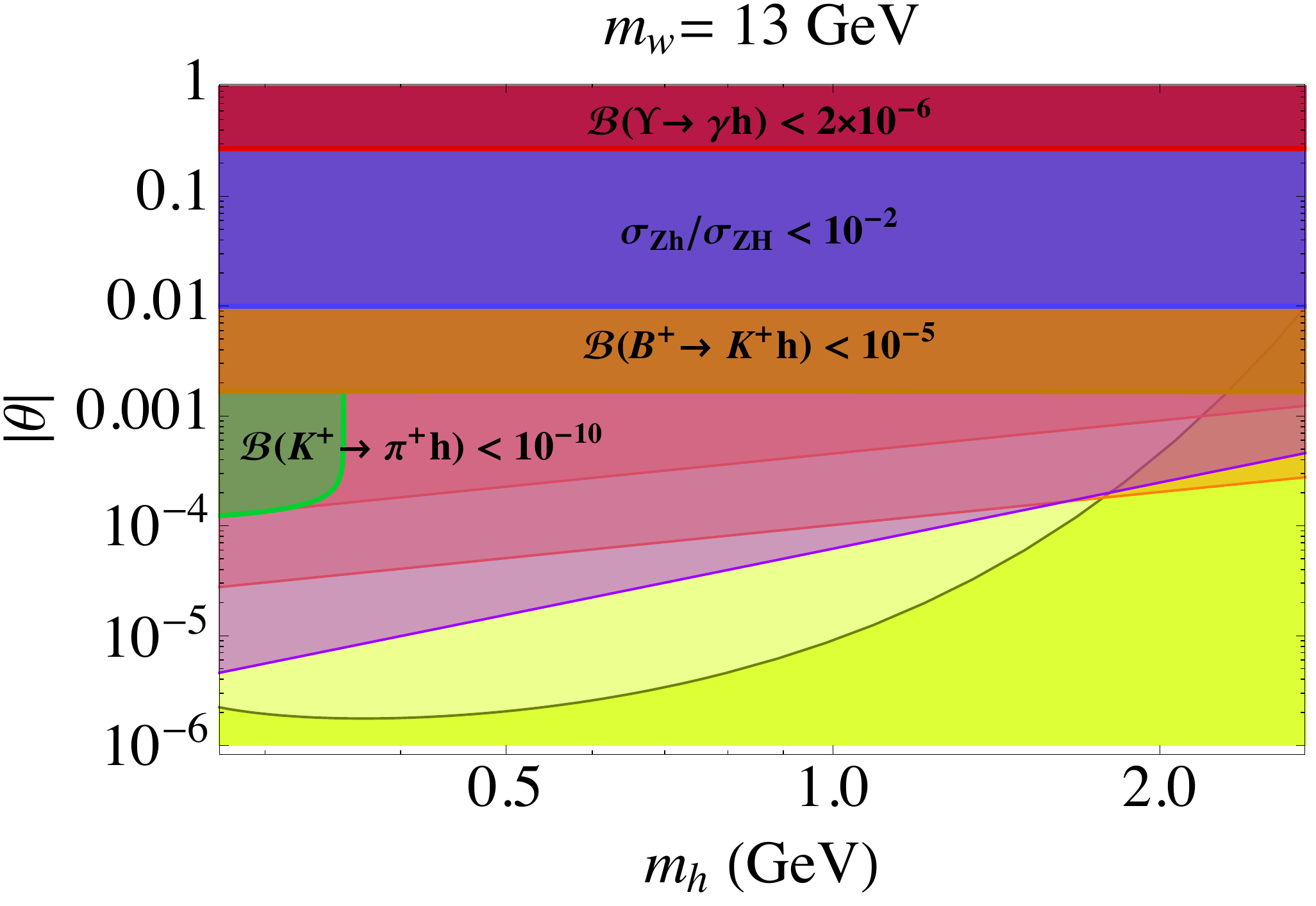}{0.99}
\end{minipage}
\caption{Bounds on the $(\theta, m_h)$ including limits on both invisible Higgs decays and direct dark matter detection for 
$m_w = 8$~Gev and $m_w = 13$~GeV.}
\label{fig:w7} 
\end{figure}

As we remarked in the Introduction, the LHC has ushered in a new era
of discovery, with confirmation of the reality of the SM
Higgs~\cite{Aad:2012tfa,Chatrchyan:2012ufa}. The ATLAS and CMS
experiments are beginning to explore in detail the properties of the
Higgs, including the various couplings to SM particles.  Since
invisible decays reduce the branching fraction to the (visible) SM
final states, it is to be expected that ${\cal B} (H \to \, {\rm
  invisible})$ is strongly constrained. Indeed ${\cal B} (H \to \,
{\rm invisible})$ is known to be less than about 19\% at
95\%CL~\cite{Espinosa:2012vu,Cheung:2013kla,Giardino:2013bma,Ellis:2013lra}.
For a Higgs width of about 4~MeV, the partial width for decay into
unobserved particles is found to be
\begin{equation}
\Gamma_{H \to \, {\rm invisible}} < 0.8~{\rm MeV} \, . 
\label{LHCwidth}
\end{equation}

Four new processes contribute to the invisible decay of the Higgs boson.
The new decay modes and the corresponding decay rates are~\cite{Anchordoqui:2013pta}
\begin{eqnarray}
\Gamma_{H \rightarrow \alpha \alpha} & = & \frac{1}{32 \pi}
\left(\frac{g_\theta \, \langle \phi \rangle }{m_H^2-m_h^2}\right)^2 m_H^3 \,,\nonumber \\
\Gamma_{H \rightarrow h h} & = & \frac{1}{32 \pi}
\left(\frac{g_\theta \, \langle \phi \rangle }{m_H^2-m_h^2}\right)^2 m_H^3 \,,\\
\Gamma_{H \rightarrow \psi_\pm \psi_\pm} & = & \frac{1}{16 \pi} \left( \frac{f
    g_\theta \, \langle r \rangle \, \langle \phi \rangle }{m_H^2 - m_h^2}\right)^2
\sqrt{m_H^2-4m_\pm^2} \ . \nonumber 
\end{eqnarray}
The decay width of the Higgs into the
hidden sector is then given by
\beq \Gamma_{H
  \rightarrow {\rm hidden}}  =  \frac{1}{16 \pi}
\left(\frac{g_\theta \, \langle \phi \rangle }{m_H^2-m_h^2}\right)^2 m_H^3 +
\frac{1}{8 \pi} \left( \frac{f \, g_\theta \ \langle r  \rangle\
 \ \langle \phi \rangle }{m_H^2 - m_h^2}\right)^2
\sqrt{m_H^2-4m_w^2} \, .
\label{eq:invis}
\eeq
Assuming $m_H \gg m_h$, this decay width is 
\beq
\Gamma_{H \rightarrow {\rm hidden}}  = \frac{g_\theta^2
  \langle \phi \rangle^2}{16 \pi m_H } + \frac{g_\theta^2 \Delta m^2 \,
  \langle\phi \rangle^2}{32
  \pi m_H^3} .
\label{eq:fifteen-1}
\eeq 
Equation~(\ref{eq:fifteen-1}) can be written in terms of the mixing angle $|\theta|$ and the quartic coupling of the hidden scalar $\lambda$,
\begin{equation}
\Gamma_{H \rightarrow {\rm hidden}} =  \frac{\theta^2 m_H}{8\pi} \left[ \lambda \frac{m_H^2}{m_h^2} +  f^2 \sqrt{1 - \frac{4 m_w^2}{m_H^2}} \, \right] \, . 
\label{uted}
\end{equation}
Equating (\ref{LHCwidth}) and (\ref{uted}) we obtain 90\% C.L. exclusion
contours in the $(|\theta|, m_h)$ plane as a function of the free
parameter $\lambda$,
\begin{equation}
|\theta (\lambda)|  <  1.27 \times 10^{-2} \ \left[ \lambda \frac{m_H^2}{m_h^2} +  f^2 \sqrt{1 - \frac{4 m_w^2}{m_H^2}} \, \right]^{-1/2} \ .
\label{fortyniners}
\end{equation}
For the region of the parameter space of interest, the second term in
(\ref{fortyniners}) is negligible.  In Fig.~\ref{fig:w6} we show the
exclusion contours for the $\lambda = 1$ and $\lambda = 0.05$.  For
smaller values of $\lambda$, bounds in the $(|\theta|, m_h)$ plane are
dominated by $B$-meson decays. Figure~\ref{fig:w7} displays the
situation including bounds from direct detection experiments for two
values of $m_w$.  One sees that for $m_w \agt 13~{\rm GeV}$ the limit
from LUX dominates the disallowed region.

\section{Conclusions}
\label{sec7}

In this article, we have examined Weinberg's Higgs portal model in
light of a variety of experimental results. In the context
of this model, we began by considering the excess of relativistic
degrees of freedom, $\Delta N_{\rm eff}$, induced by the weakly
interacting Goldstone bosons, $\alpha'$, which decouple from SM
particles in the late early universe. For masses of the hidden scalar
in the range $0.2~{\rm GeV} \alt m_h \alt 4~{\rm GeV}$, the
interaction rate of Goldstone bosons is dominated by resonant
annihilation into fermion-antifermion pairs. In a recent
calculation~\cite{Garcia-Cely:2013nin} this thermal annihilation rate
was derived using Maxwell-Boltzmann statistics. We have verified with
a full expansion of the Bose-Einstein distribution that the leading
term provides the results of~\cite{Garcia-Cely:2013nin}, as well as
negligible higher order terms. The decoupling temperature,
$T_{\alpha'}^{\rm dec}$, determines features of the contours in
$(|\theta|, m_h)$ parameter space, where $\theta$ is the mixing angle in the Higgs sector. Following~\cite{Weinberg:2013kea} we take as fiducial the
contour for $T_{\alpha'}^{\rm dec} = m_\mu$ corresponding to $N_{\rm
  eff} = 3.39$, which is a compromise between the number of effective
neutrino species emerging from multi-parameter fits to Planck data in
which the value $h$ is, in one case, allowed to float in the fit, and
in the second case is frozen to the value determined by the Hubble
Space Telescope. This contour divides the parameter space into a lower
region consistent with SM physics at the 1$\sigma$ level and an upper
region requiring new physics at sub-fermi distance.

We then proceed to constrain the new physics regime using data from a
variety of sources.  First, we use data from
BaBar~\cite{delAmoSanchez:2010bk,Lees:2013kla,Aubert:2004ws},
CLEO~\cite{Browder:2000qr}, BELLE~\cite{Lutz:2013ftz},
E787~\cite{Adler:2001xv}, and
E949~\cite{Anisimovsky:2004hr,Adler:2008zza,Artamonov:2009sz} on
decays of heavy mesons with missing energy. Using results from
searches for \mbox{$B^+ \to K^+ + \met$} we derived an upper limit on
the mixing angle that improves by one order of magnitude the latest
bound derived in~\cite{Cheung:2013oya} from LEP limits on the
production of invisibly-decaying Higgs
bosons~\cite{Barate:1999uc,Abdallah:2003ry,Achard:2004cf,Abbiendi:2007ac}.
For $m_h \alt 355~{\rm MeV}$, measurements of $K^+ \to \pi^+ + \met$
further improve the upper limit on $|\theta|$ by up to two orders of
magnitude.  The bounds resulting from our analysis rule out a
significant part of the parameter space favored by
CoGeNT~\cite{Aalseth:2010vx,Aalseth:2011wp,Aalseth:2012if} and CDMS
II~\cite{Agnese:2013rvf}. They are complementary (and comparable) to:
{\it (i)}~recent results from the ATLAS
Collaboration~\cite{Aad:2014iia} yielding bounds on a Higgs portal
model in which the only interaction with the fermionic dark matter is
through the SM Higgs~\cite{Kanemura:2010sh,Fox:2011pm,Djouadi:2011aa};
{\it (ii)}~bounds established from searches for $B$ meson decay into
charged leptons which have been used to constrain light scalar
couplings in other Higgs portal models~\cite{Schmidt-Hoberg:2013hba}.

Next, we considered the implication of measurements from DM direct
detection experiments. We extended the analysis developed
in~\cite{Anchordoqui:2013pta} to incorporate new reults from 
 CDMSlite~\cite{Akerib:2013tjd} and
 LUX~\cite{Akerib:2013tjd}. We found exclusion regions on the $(|\theta|, m_h)$ plane
which are in agreement with those recently reported
in~\cite{Garcia-Cely:2013nin}.  We have shown that for light WIMPs,
$m_w \alt 9$~GeV, bounds from meson decay are more restrictive than
limits from DM direct detection experiments. This is an interesting
region to scrutinize, as signals have been reported both
before~\cite{Aalseth:2010vx,Aalseth:2011wp,Aalseth:2012if,Agnese:2013rvf} and after~\cite{Aalseth:2014eft} the LUX
bounds were published~\cite{Akerib:2013tjd}. On the other hand, LUX measurements exclude
heavy WIMPs, $m_w \agt 100$~GeV.  For $9~{\rm GeV} \alt m_w \alt
10~{\rm GeV}$, the best limits come from CDMSlite, while in the
intermediate regions from $10~{\rm GeV} \alt m_w \alt 55~{\rm GeV}$
and $70~{\rm GeV} \alt m_w \alt 100~{\rm GeV}$, LUX provides the best
bounds. (In the region from about 55~GeV to 70~GeV the present analysis
is not valid due to the presence of the Higgs resonance).

Finally, we considered constraints from LHC searches for invisible
Higgs decays~\cite{Espinosa:2012vu,Cheung:2013kla,Giardino:2013bma,Ellis:2013lra} for different assumptions about the Higgs quartic
coupling, $\lambda$. These measurements further constrain the
parameter space for $m_w \alt 13$~GeV if the $\lambda \agt 0.05$.  

In summary, we have compressed the allowed parameter space for
Weinberg's Higgs portal model using a variety of complementary
methods. Future measurements will deeply probe this model.
Measurements by ATLAS and CMS with LHC running at 14~TeV
center-of-mass energy will measure the Higgs couplings, and in
conjunction with measurements from LHCb and NA62~\cite{Moulson:2013oga}, will provide a window to the
low mass WIMP.  Direct detection experiments like LUX and XENON1T~\cite{Aprile:2012zx}
will either detect or constrain WIMPs with masses in the 10's of
GeV.

\section*{ Note Added}

Shortly after this paper was written the SuperCDMS Collaboration
presented new bounds on the low mass WIMP spin-independent interaction
cross section~\cite{Agnese:2014aze}. These bounds, when translated into excluding
regions of ($\theta, m_h$) plane, provide the most restrictive limits
on the mixing angle for $8~{\rm GeV} \alt  m_w \alt 9~{\rm GeV}$.

\section*{Acknowledgments}
This work was supported in part by the US NSF grants:
CAREER PHY1053663 (LAA), PHY-0757959 (HG), PHY-1205854 (TCP),
US DoE grant DE-FG05-85ER40226 (TJW), NASA NNX13AH52G (LAA, TCP), 
UWM RGI (BJV), and UWM Physics 2014 Summer Research Award (LHMdaS).

\subsection*{Appendix A}

Consider the annihilation process
$\alpha(k) \ \alpha(k') \rightarrow \bar {\rm f}(p) \ {\rm f}(p')$ in the center-of-mass frame, with
the initial 3-momentum of the
$\alpha$ particles given as ${\bf k} = -{\bf k'}$, respectively.  The
differential cross section is given by 
\begin{equation} 
 d \sigma = \frac{1}{2}
\frac{\overline{|\CM|}^2}{2 k 2 k'} \frac{d^3 p}{(2\pi)^2 2 E_p}
\frac{d^3 p'}{2 E_{p'}} \delta^{(3)}( {\bf p} + {\bf p'}) \delta(2k -
E_p - E_{p'}) \ .  
\end{equation} 
Use of the spatial delta function allows us to write this result as
\begin{eqnarray}
d \sigma & = & \frac{\overline{|\CM|}^2}{8 k^2} \frac{d^3 p}{(2\pi)^2
  4E_p^2}  \frac{1}{2} \delta (k -E_p) \nonumber \\
 & =  &\frac{\overline{|\CM|}^2}{8  k^2 } \frac{p dE_p
   d\Omega}{(2\pi)^2 4 \,E_p  } \ \frac{1}{2} \delta \left(E_p -k \right) \ .
\end{eqnarray}
After performing the integration over $E_p$ we arrive at
\beq  d \sigma = \frac{1}{2}\frac{\overline{|\CM|}^2}{32 \pi k^2}
\frac{d\Omega}{4 \pi} \frac{\sqrt{k^2- m_{\rm f}^2} }{k}\ .  \eeq 
In terms of the invariant Mandelstam variable, $s= ( k + k')^2 = 4 k^2$, we can write the differential cross section as
\beq  d \sigma =
\frac{\overline{|\CM|}^2}{16 \pi} \frac{d\Omega}{4 \pi}
\frac{\sqrt{s-4m_{\rm f}^2} }{s^{3/2}} \ .
\label{previous}
\eeq

We now turn to evaluate the invariant scattering amplitude.  The
Lagrangian describing the interaction of the Goldstone bosons with SM
fields contains the Yukawa terms~\cite{Anchordoqui:2013pta} \beq
\frac{m_{\rm f}}{\langle \phi \rangle} H \bar {\rm f} {\rm f} \cos
\theta - \frac{m_{\rm f}}{\langle \phi \rangle} h \bar{\rm f} {\rm f}
\sin \theta \, ,
\label{lapicera1}
\eeq and the terms coupling the Goldstone bosons with Higgs doublet and
the $CP$-even scalar  \beq \frac{\sin \theta}{\langle r \rangle} H
(\p \alpha)^2 + \frac{\cos \theta}{\langle r \rangle} h (\p \alpha)^2
\ .  
\label{lapicera2}
\eeq From (\ref{lapicera1}) and (\ref{lapicera2}) we get the Feynman
rules for: the $(\alpha, \alpha, H)$ vertex, \mbox{$ -i 2 \sin \theta
  \, (k \cdot k') /\langle r \rangle$}; the $(\alpha, \alpha, h)$
vertex, $ -i\, 2 \cos \theta \, (k \cdot k')/\langle r \rangle$; the
$(H, \bar {\rm f}, {\rm f})$ vertex, $i \, m_{\rm f} \cos
\theta/\langle \phi \rangle$; and the $(h, \bar {\rm f}, {\rm f})$
vertex, $ i \, m_{\rm f} \sin \theta / \langle \phi \rangle$. All in
all, the $s$-channel invariant amplitude of the process mediated by
$H$ and $h$ can be expressed as \beq \CM = \frac{2 \sin \theta \cos
  \theta}{\langle r \rangle \, \langle \phi \rangle} (k \cdot k')
\left(\frac{m_H^2 -m_h^2}{(s-m_H^2)(s-m_h^2)} \right) \bar u(p') v(p)
\ .  \eeq Hence, for $\theta \ll 1$, the spin summed-average
square amplitude is found to be \beq \overline{|\CM|}^2 = \frac{4 \,
  m_{\rm f}^2 \, g_\theta^2}{(s-m_H^2)^2(s-m_h^2)^2} (k \cdot k')^2
4(p \cdot p' - m_{\rm f}^2) \,, \eeq or in terms of invariant
variables \beq \overline{|\CM|}^2 = \frac{2\, m_{\rm f}^2 \,
  g_\theta^2}{(s-m_H^2)^2(s-m_h^2)^2} s^2 (s- 4m_{\rm f}^2) \ .
\label{ma8}
\eeq Substituting (\ref{ma8}) into (\ref{previous}) we obtain \beq
\sigma_{\alpha \alpha} = \frac{4}{32 \pi} \frac{m_{\rm f}^2
  g_\theta^2}{(s-m_H^2)^2(s-m_h^2)^2} s^2 \left(1- \frac{4m_{\rm
      f}^2}{s} \right)^{3/2} \ .  \eeq 
For phenomenological purposes,
the poles need to be softened to a Breit-Wigner form by obtaining and
utilizing the correct total widths of the resonances; {\it e.g.}, for $\sqrt{s} \sim m_h$,
\beq
\frac{i}{s-m_h^2} \rightarrow \frac{i}{s - m_h^2 + i m_h \Gamma_h} \ .
\eeq
After this is done, the scattering cross section becomes
\beq
\sigma_{\alpha \alpha}  = \frac{4}{32 \pi}  \frac{m_{\rm f}^2 \, \theta^2}{\langle r \rangle^2 \, \langle \phi \rangle^2}\frac{(m_H^2 - m_h^2)^2 + m_h^2 \Gamma_h^2}{(s-m_H^2)^2 \, \left[(s-m_h^2)^2 + m_h^2 \Gamma_h^2\right]} s^2 \left(1-\frac{4m_{\rm f}^2}{s} \right)^{3/2} \ .
\eeq
For $m_h^2 \Gamma_h^2 \ll (m_H^2-m_h^2)^2$, we can drop the $\Gamma_h$ term in the numerator to obtain
\beq
\sigma_{\alpha \alpha} = \frac{4}{32 \pi}  \frac{m_{\rm f}^2 \, g_\theta^2}{(s-m_H^2)^2 \, \left[(s-m_h^2)^2 + m_h^2 \Gamma_h^2\right]} s^2 \left(1-\frac{4m_{\rm f}^2}{s} \right)^{3/2} \ .
\label{sesentayuno}
\eeq
We have found that for the considerations in the present work, the cross section can be safely approximated by the single pole of the Narrow-Width Approximation. Namely, for $m_h \Gamma_h \rightarrow 0$, (\ref{sesentayuno}) can be rewritten as
\beq
\sigma_{\alpha \alpha} \approx \frac{4}{32}  \frac{m_{\rm f}^2 g_\theta^2}{(s-m_H^2)^2} s^2 \left(1-\frac{4m_{\rm f}^2}{s} \right)^{3/2}   \frac{\delta(s-m_h^2)}{m_h \Gamma_h} \ ,
\eeq
where we have used the relation 
\beq
\lim_{\epsilon \rightarrow 0} \frac{\epsilon}{x^2 + \epsilon^2} = \pi \delta(x) \ .
\eeq

\begin{figure}[h]
\begin{center}
\postscript{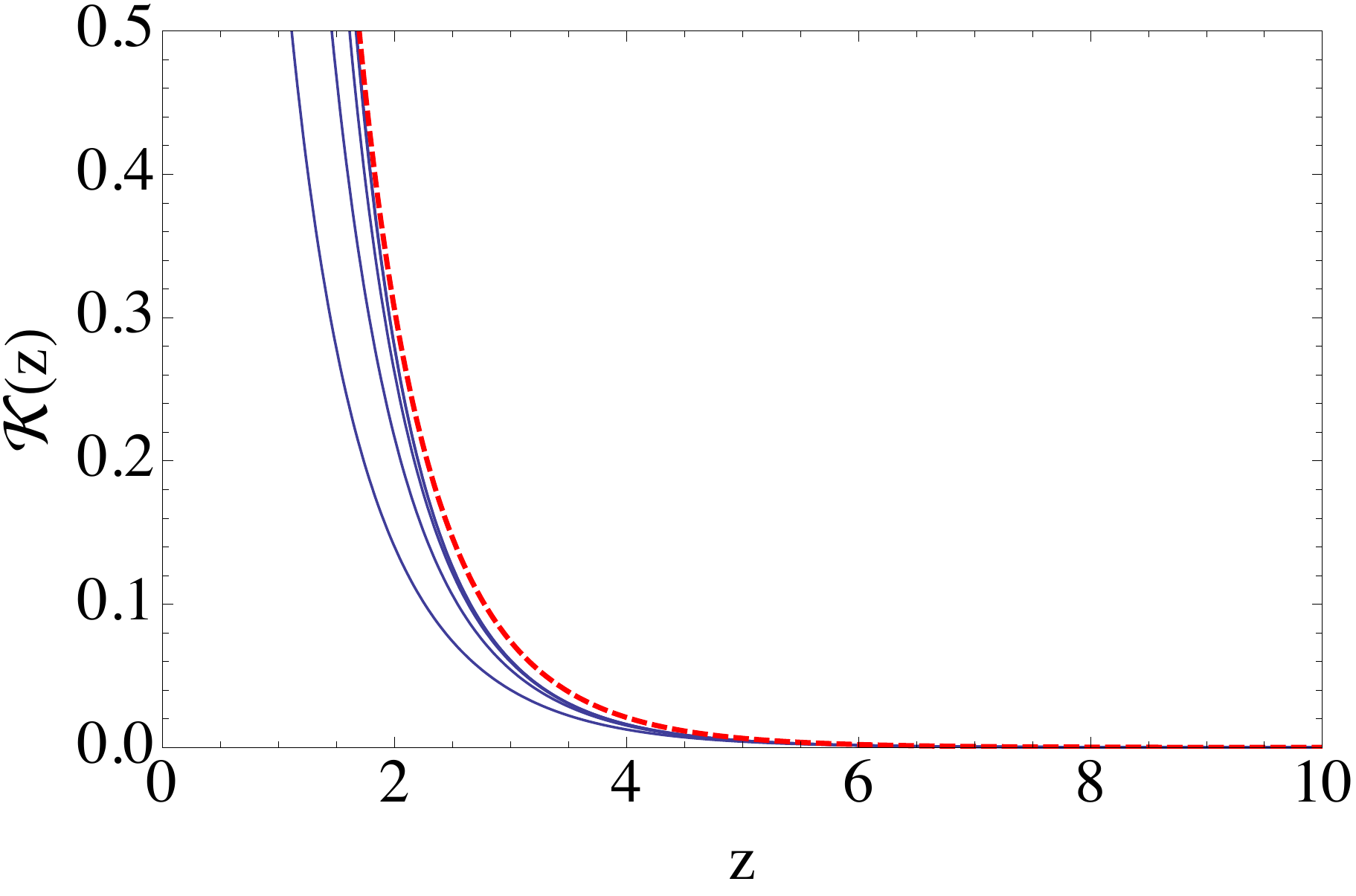}{0.7}
\caption{The solid lines stand for ${\cal K} (z) = \sum_{n,k=0}^N  K_1( \sqrt{(k+1)(n+1)} z)/\sqrt{(k+1)(n+1)}$ as a function of $z = m_h/T$.  From left to right $N = 0, 1, 3, 10, 50$. The sum quickly converges towards $\zeta(3) K_2(z)$, which is shown as a dashed curve.}
\label{fig:w8} 
\end{center}
\end{figure}

We now proceed to calculate the thermal-angular averages,
\beq
\la \sigma_{\alpha \alpha} v_M \ra = \frac{1}{n^2_\alpha} \int \frac{d^3 p_1}{(2\pi)^3}\frac{d^3 p_2}{(2\pi)^3}  \sigma_{\alpha \alpha} v_M (s) \ \frac{1}{e^{E_1/T} - 1}\frac{1}{e^{E_2/T} - 1} \ ,
\label{sopaotravez}
\eeq
where $v_M$ is the is the M\"oller velocity~\cite{Weiler:2013hh}.
Substituting the expansion 
of the Bose distribution,
\beq
\frac{1}{e^{E/T}-1} = e^{-E/T} \sum_{n=0}^\infty e^{-n E/T} \,, 
\label{eq:expand}
\eeq
into (\ref{sopaotravez}) we obtain~\cite{Gondolo:1990dk}
\beq
\la \sigma_{\alpha \alpha} v_M \ra = \frac{T}{8 \pi^4 n^2_\alpha(T)} \sum_{k,n=0}^{\infty}  \frac{1}{4 \sqrt{(k+1)(n+1)}} \int_{0}^\infty \sigma_{\alpha \alpha} \, s \sqrt{s} K_1\left( \frac{\sqrt{(k+1)(n+1) s}}{T} \right) ds\,, \nonumber
\eeq
with 
\beq
n_\alpha = \int_0^\infty \frac{d^3 p}{(2\pi)^3} \frac{1}{e^{p/T}-1} =  \frac{\zeta(3)}{\pi^2}T^3 \ .
\eeq
Using the narrow width approximation to the annihilation cross section we have
\begin{eqnarray}
\la \sigma_{\alpha\alpha} v_M \ra & = & \frac{4 m_{\rm f}^2 g_\theta^2 }{32^2 \zeta^2(3) T^5 \Gamma_h} \sum_{n,k=0}^\infty \frac{1}{\sqrt{(k+1)(n+1)}} \frac{m_h^6}{(m_h^2-m_H^2)^2} \left(1- \frac{4 m_{\rm f}^2}{m_h^2} \right)^{3/2} \nonumber \\
& \times & K_1\left( \frac{\sqrt{(k+1)(n+1)} m_h}{T} \right) \ ,
\end{eqnarray} 
which under the assumption that $m_h \ll m_H$ results in
\begin{eqnarray}
\la \sigma_{\alpha \alpha} v_M \ra & = & \frac{g_\theta^2}{256} \frac{m_{\rm f}^2 m_h^6 }{\zeta^2(3) T^5 m_H^4 \Gamma_h} \left(1- \frac{4 m_{\rm f}^2}{m_h^2} \right)^{3/2} \sum_{n,k=0}^\infty \frac{1}{\sqrt{(k+1)(n+1)}} \nonumber \\
& \times & K_1\left( \frac{\sqrt{(k+1)(n+1)} m_h}{T} \right) \ .
\label{anteU}
\end{eqnarray}
If we retain only the first term in the series we recover the result obatined in~\cite{Garcia-Cely:2013nin} using Maxwell-Boltzmann statistics, 
\begin{equation}
\la \sigma_{\alpha \alpha} v_M \ra  \approx  \frac{g_\theta^2}{256} \frac{m_{\rm f}^2 m_h^6 }{ \zeta^2(3) T^5 m_H^4 \Gamma_h} \left(1- \frac{4 m_{\rm f}^2}{m_h^2} \right)^{3/2}  K_1(m_h/T) \ .
\end{equation}
To check the accuracy of retaining only the first term of the series, we perform  a numerical integration, which is shown in Fig.~\ref{fig:w8}.
The summation converges within the first $N=10$ terms of the
series. For $T \alt m_h$, we find agreement at better than the 7\%
level between the $2^{\rm nd}$ Bessel function of the second kind and
the double series, \beq \sum_{n,k=0}^\infty
\frac{1}{\sqrt{(k+1)(n+1)}} K_1\left( \frac{\sqrt{(k+1)(n+1)} m_h}{T}
\right) \approx \zeta(3) K_2(m_h/T) \,, \eeq as demonstrated in
Fig.~\ref{fig:w8}. This allows us to write an approximate expression
for the double series and thus (\ref{anteU}) becomes, \beq \la
\sigma_{\alpha \alpha} v_M \ra \approx \frac{g_\theta^2}{256}
\frac{m_{\rm f}^2 m_h^6 }{ \zeta(3) T^5 m_H^4 \Gamma_h} \left(1-
  \frac{4 m_{\rm f}^2}{m_h^2} \right)^{3/2} K_2(m_h/T) \ .
\label{sevalaU}
\eeq
Multiplying (\ref{sevalaU}) by the number density $n_\alpha$ we obtain the interaction
rate given in (\ref{XsX}).

\section*{Appendix B}

For completness, we briefly recall here the basics of the calculation
of the $w$ relic density~\cite{Kolb:1990vq}. The evolution of the number desnity
$n_w$ is governed by the Boltzmann transport equation \beq \dot n_w +
3 H n_w = -\la \sigma_{ww} v_M \ra (n_w^2 - n_{w_{\rm EQ}}^2) \,, 
\label{boltzmannEQ}
\eeq
where $x = m_w/T$. In the non-relativistic limit, and in the
Maxwell-Boltzmann approximation, the number density at thermal
equilibrium is given by \beq n_{w_{\rm EQ}} = g(x) \left( \frac{m_w^2 }{2
    \pi} \right)^{3/2} x^{-3/2} e^{-x} \, . \eeq We next introduce the
yield variable $Y \equiv n_w/s$, where
\begin{equation}
s = g (x) \, \frac{2 \pi^2}{45} \, m_w^3 \, x^{-3}
\end{equation}
is the entropy density.\footnote{If relativistic particles are present
  that have decoupled from the plasma, it is necessary to distinguish
  between two kinds of $g$: $g_\rho$ which is associated with the total energy
  density, and $g_s$ which is associated with the total entropy
  density. For our calculations we use $g = g_\rho = g_s$.} In particular,
\begin{equation}
Y_{\rm EQ} \equiv \frac{n_{w_{\rm}
    EQ}}{s} = \frac{45}{\sqrt{32 \, \pi^7}} \, x^{3/2} \ e^{-x} \, . 
\label{yieldEQ}
\end{equation}
Using the conservation of entropy per comoving volume, it follows that 
\beq
 \dot s + 3 H s =0 \,,
\end{equation}
or equivalently
\begin{equation}
\dot x =  H(x) x, 
\end{equation}
where
\begin{equation}
\ H(x) = \sqrt{\frac{8\pi^3}{90}}  \frac{1}{M_{\rm Pl}} \, \sqrt{g(x)} \, m_w^2 \, x^{-2} = H(m_w) \, x^{-2} \ .
\end{equation}
Equation (\ref{boltzmannEQ}) can now be expressed in terms of $x$ and $Y$ variables to obtain
\beq
\frac{dY}{dx}  = -\frac{x \la \sigma_{ww} \, v_M \ra \, s}{H(m_w)} \, (Y^2-  Y_{\rm EQ}^2 ) \ ,
\label{byby}
\eeq 
where 
\beq
\frac{x \, \la \sigma_{ww} v_M \ra \, s}{H(m_w)} =\sqrt{\frac{\pi}{45}} \, \sqrt{g(x)} \, m_w \, M_{\rm Pl} \, \frac{\la \sigma_{ww} v_M \ra \ }{x^2}.
\eeq
After reparametrizing the yield, $Y = Y_{\rm EQ}+ \Delta$, in (\ref{byby})  we obtain 
\begin{equation} 
\frac{dY_{\rm EQ}}{dx} + \frac{d \Delta}{dx} = -\sqrt{\frac{\pi}{45}} \sqrt{g (x)} \, m_w \, M_{\rm Pl}  \frac{ \la \sigma_{ww} v_M \ra}{x^2} \Delta \left(  2 Y_{\rm EQ} + \Delta \right) \ .
\label{pingot}
\end{equation}
Near freeze out $d\Delta/dx \approx 0$, and so (\ref{pingot}) simplifies to
\begin{equation}
\frac{1}{Y_{\rm EQ}} \frac{dY_{\rm EQ}}{dx} \approx -\sqrt{\frac{\pi}{45}} \sqrt{g(x_f)} \,  m_w \,  M_{\rm Pl} \, \frac{ \la \sigma_{ww} v_M \ra}{x_f^2} \ c \, (c+2)   \, Y_{\rm EQ} \,,
\label{CGPF}
\end{equation}
where we have taken $\Delta (x_F) = c Y_{\rm EQ} (x_f)$ to define the
freeze out time. Here, $c$ is a constant of order one determined by
matching the late-time and early-time solutions. Finally substitution
of (\ref{yieldEQ}) into (\ref{CGPF}) leads to
\begin{equation}
e^{x_f}  \approx
c \, (c+2) \, \sqrt{\frac{45}{32 \pi^6}} \sqrt{g (x_f)} \, m_w  \,M_{\rm Pl} \frac{\la \sigma_{ww} v_M \ra}{x_f^{1/2}}  \,. 
\label{iterative}
\end{equation} 
The freeze-out temperature $x_f$ can be estimated through the
iterative solution of (\ref{iterative}), yielding
\beqa
x_f &\approx& \ln \left[ c (c+2)  \sqrt{\frac{45}{32 \pi^6}} \,
  \sqrt{g(x_f)} \,  m_w \, M_{\rm Pl} \la \sigma_{ww} v_M \ra \right]
- \frac{1}{2} \ln x_f   \nonumber \\
&\approx& \ln \left[0.1 \, \sqrt{g(x_f)} \,  m_w \,  M_{\rm Pl} \, \la \sigma_{ww} v_M \ra \right] - \frac{1}{2} \ln \left( \ln \left[ 0.1 \, \sqrt{g(x_f)} m_w \, M_{\rm Pl} \la \sigma_{ww} v_M \ra \right] \right) . 
\label{eq:A1}
\eeqa After  freeze out the yield  significantly departs  from
its equilibrium expression. Thus, to obtain the $Y$ evolution for $x
\gg x_f$, we can neglect the 
$Y_{\rm EQ}$ terms in (\ref{pingot}) as the $\Delta$ terms come to dominate, \beq
\frac{d \Delta}{dx} = -\sqrt{\frac{\pi}{45}} 
  \sqrt{g (x)} \,  m_w \, M_{\rm Pl} \, \frac{ \la \sigma_{ww} v_M \ra}{x^2} \Delta^2 \
.  \eeq Upon solving for $\Delta$ at todays value, $x_0$, we obtain 
 \beq \frac{1}{Y(x_0)} =
\frac{1}{\Delta(x_f)}+ \sqrt{\frac{\pi}{45}} \, m_w M_{\rm Pl} \int_{x_f}^{x_0}
\sqrt{g (x)} \frac{ \la \sigma_{ww} v_M \ra}{x^2} dx \ ,
\label{thrhs}
\eeq  
where we have taken $\Delta(x_0) \approx Y( x_0)$.  Assuming that
$g(x)$ remains roughly constant over  the integration range $(x_f, x_0)$ and
that $\la \sigma_{ww} v_M \ra  \propto x^{-n}$, the first term in the
right-hand-side of (\ref{thrhs}) becomes
\begin{equation}
\Delta^{-1}(x_f)  = (c+2) \sqrt{\frac{\pi}{45}} m_w M_{\rm Pl}  \sqrt{g(x_f)} \frac{ \la \sigma_{ww} v_M(x_f) \ra}{x_f^2} \ , 
\end{equation}
whereas the second term is given by
\begin{equation}
 \sqrt{\frac{\pi}{45}}\, m_w \ M_{\rm Pl} \sqrt{g(x_f)} \int_{x_f}^{x_0}\frac{ \la \sigma_{ww} v_M \ra}{x^2} dx \approx \sqrt{\frac{\pi}{45}} m_w M_{\rm Pl}  \sqrt{g(x_f)} \frac{\la \sigma_{ww} v_M(x_f) \ra}{(n+1) x_f} \ .
\end{equation}
Since the first term $\propto  \la \sigma_{ww} v_M(x_f)
\ra/ x_f^2$ and the second term  $\propto  \la \sigma_{ww} v_M(x_f)
\ra/ x_f$, for simplicity  herein we neglect the $\Delta^{-1}(x_f)$
contribution to (\ref{thrhs}). The present density of  $w$ is simply given by $\rho_w = m_wn_w = m_w
s_0 Y (x_0)$, where  $s_0 = 2890.7(9) \ {\rm cm^{-3}} = 
2.2211(4) \times 10^{-38} \ {\rm GeV^3}$ is  the present entropy
density (assuming three Dirac neutrino species)~\cite{Beringer:1900zz}. The relic density can
finally be expressed in terms of the critical density
\beq \Omega_{\rm DM} h^2 = \frac{8 \pi m_w s_0 Y(x_0)}{3
  M_{\rm Pl}^2 (100 \ {\rm km/s/Mpc})^2} =  2.74 \times 10^8 \ {\rm
    GeV^{-1}} \,  m_w  \, Y(x_0) \ .  
\label{kilombo}
\eeq
Substituting (\ref{thrhs}) into (\ref{kilombo}) we then have
\beq \la \sigma_{ww} v_M(x_f) \ra= (n+1) \frac{
 1.04 \times 10^9 \ {\rm GeV^{-1}} \, x_f}{\sqrt{g(x_f)} \, 
  M_{\rm Pl} \, \Omega_{\rm DM} \, h^2} \ ,  \eeq
where we have taken $T_0 = 2.7255~{\rm K}$~\cite{Beringer:1900zz}.
For $\Delta m/m_w \to 0$, we have $n = 0$ leading to (\ref{eq:Dec3}).

\end{document}